\documentclass[%
 reprint,
nofootinbib,
 amsmath,amssymb,
 aps,
]{revtex4-2}
\usepackage{CJK}
\usepackage{graphicx}
\usepackage{dcolumn}
\usepackage{bm}
\usepackage{subcaption}
\usepackage{float}
\usepackage{amsmath}
\usepackage{cleveref}
\DeclareMathOperator{\arctanh}{\mathrm{arctanh}}

\begin{document}
\begin{CJK*}{UTF8}{gbsn}
\title{Exact four-vector work distribution and covariant fluctuation theorems of work for a relativistic particle in an expanding piston}

\author{Tingzhang Shi (石霆章)}
\thanks{Equal contribution}
\affiliation{
 School of Physics, Peking University, Beijing, 100871, China
}

\author{Chentong Qi (齐晨桐)}
\thanks{Equal contribution}
\affiliation{
 School of Physics, Peking University, Beijing, 100871, China
}

\author{H. T. Quan (全海涛)}\email{htquan@pku.edu.cn}
\affiliation{
 School of Physics, Peking University, Beijing, 100871, China
}
\affiliation{
 Collaborative Innovation Center of Quantum Matter, Beijing 100871, China
}
\affiliation{Frontiers Science Center for Nano-optoelectronics, Peking University, Beijing, 100871, China}

\date{\today}

\begin{abstract}
We investigate the non-equilibrium four-vector work in an expanding relativistic piston. 
We derive the exact work distribution in this pedagogical model and find that the joint distribution of four-vector work $(W^0, W^1)$ concentrates on the origin and some curves in the $(W^0, W^1)$ space, rather than being smoothly distributed.
In the non-relativistic limit, our model consistently recovers the non-relativistic dynamics.
We further demonstrate that the momentum component of four-vector work remains significant in both the Lorentz-relativistic and Galilean-relativistic frameworks.
On top of the work distribution, we verify a family of covariant fluctuation theorems of work.
In addition, we introduce a novel geometrical technique for analyzing the dynamics of relativistic collision processes, which can be straightforwardly extended to multi-dimensional piston models.
\end{abstract}

\maketitle
\end{CJK*}

\section{\label{sec:introduction}Introduction}
Covariance is the crucial requirement for a physical theory.
However, the unification of special relativity and thermodynamics has been a longstanding conundrum \cite{hanggi09}.
On the one hand, the covariant generalizations of the laws of thermodynamics are not so straightforward \cite{kampen68,kampen69,israel79,israel81}.
Some thermodynamic relations, such as the Carnot theorem, have not been generalized to a covariant form.
Fortunately, stochastic thermodynamics provides a new perspective on the covariance principle in thermodynamics.
Unlike the second law of thermodynamics or the Carnot theorem, which are inequalities, the fluctuation theorems (FTs), which are a family of equalities valid for arbitrary far-from-equilibrium processes \cite{jar97, crooks99,martin08,parrondo07, quan23}, could be generalized to the covariant forms without ambiguity \cite{jihui25}.
On the other hand, the lack of exactly solvable models, due to the absence of action at a distance in special relativity, makes it difficult to verify the theories of relativistic thermodynamics explicitly.
One cannot freely choose a time-dependent potential to perform work on a particle (as we usually do in the non-relativistic stochastic thermodynamics).
Luckily, the elastic piston model \cite{lua05,xianghang24,quan12,zhongping14,fei19,qiu20,gong16} is an ideal choice for relativistic dynamics.
The elastic collision is a purely local interaction, and we do not need to introduce any fields to characterize the process of interaction, which significantly simplifies the details of the dynamics.

From a more fundamental perspective, fluctuation theorems (FTs) \cite{jar97, crooks99,martin08,parrondo07, quan23} provide a more robust starting point. As a family of exact equalities, FTs reconcile the time-reversible nature of microscopic dynamics with the irreversibility of macroscopic phenomena. Recently, these relations have been elevated to a manifestly covariant form without ambiguity \cite{jihui25}. This development offers a new viewpoint on the covariance principle in thermodynamics and suggests that relativistic thermodynamics may be more naturally formulated in terms of fluctuation theorems rather than entropy inequalities.

In a previous work \cite{xianghang24}, we (two of us and a collaborator) have generalized the piston model in Ref.~\cite{lua05} to a special relativistic setting, and have verified the Jarzynski equality (JE) \cite{jar97} in relativistic regime.
In the low speed and low temperature limit, we have recovered the results in Ref.~\cite{lua05}.
However, the observer is fixed in the laboratory frame of reference.
Hence, both the work distribution and the JE do not satisfy the principle of covariance \cite{lua05,xianghang24}.

In this article, in order to investigate the fluctuation theorems in a covariant form, we consider an arbitrary inertial observer.
Utilizing a novel orthogonal reflection method, we map the motion of a particle in the piston into the motion of a particle in free space, then we prove the covariant Liouville's theorem and the detailed fluctuation theorem in the one-dimensional relativistic piston model.
We analytically compute the four-vector work distribution and verify the covariant Crooks fluctuation theorem \cite{crooks99} and Jarzynski equality (JE) \cite{jar97,jihui25} in the piston model. 
In this relativistic covariant setting, we need to consider the joint distribution of both the energy and momentum change of the particle due to the elastic collision with the two boundaries of the cylinder.
We find that although the distributions of four-vector work for different inertial observers are different, they all satisfy the same covariant FTs~\cite{jihui25}. 
Hence, we demonstrate the covariance principle of fluctuation theorem in this pedagogical model.

The article is organized as follows: 
In Sec.~\ref{sec:model}, we introduce a novel coordinate transformation technique to calculate the trajectory of the particle.
In Sec.~\ref{sec:work}, we analytically calculate the four-vector work distribution of a relativistic particle in an expanding piston system perceived by an arbitrary inertial observer.
We also find that the joint distribution of the four-vector work $W^\mu=(W^0, W^1)$ concentrates on the origin and some curves in the $(W^0, W^1)$ space.
In Sec.~\ref{section:hs}, we demonstrate the validity of covariant FTs of work in the one-dimensional relativistic piston model.
In Sec.~\ref{sec:dis}, we give the discussion and summary.

\section{\label{sec:model}A Geometric Method For Solving the Relativistic Piston Model}
\begin{figure}
    \centering
    \includegraphics[width=0.94\linewidth]{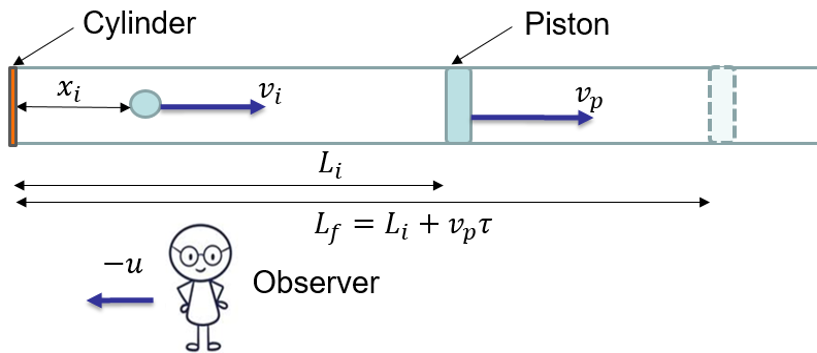}
    \captionsetup{justification=raggedright,singlelinecheck=false}
    \caption{A relativistic particle moving in a one-dimensional cylinder. The left boundary is fixed in the laboratory reference frame, while the right boundary is a elastic piston with constant velocity $v_p$. An inertial observer is moving relative to the laboratory frame with a constant velocity $-u$.
    The initial position and velocity of the particle are $x_i$ and $v_i$, respectively.}
    \label{fig:cartoon demo}
\end{figure}

The model we consider here is a single classical relativistic particle inside a one-dimensional cylinder \cite{lua05,xianghang24,quan12,zhongping14,qiu20,gong16,nolte09}, see FIG.~\ref{fig:cartoon demo}.
The mass of the particle $m$ is set to be $m=1$.
In the cylinder's frame, the initial length of the vessel is $L_i$, and the gas is initially at the inverse temperature $\beta$, where $\beta = 1/(k_{B} T)$, $k_B$ is the Boltzmann constant, and $T$ is the temperature perceived from the cylinder's frame.
We now fix the left boundary and expand the right boundary outwards at the speed $v_p$, and stop it after a time interval $\tau$. The final length of the vessel perceived from the cylinder's frame is thus $L_f=L_i+v_p \tau$.
In the following discussion, we denote the coordinate of the left boundary as the coordinate of the cylinder, and the coordinate of the moving boundary on the right as the coordinate of the piston.
The particle undergoes several elastic collisions with the piston and the cylinder, which means that in the piston's frame, during a collision, the energy of the particle is a constant, and the direction of the momentum is reversed. 

\begin{figure}[ht]
    \centering
    \includegraphics[width=0.95\linewidth]{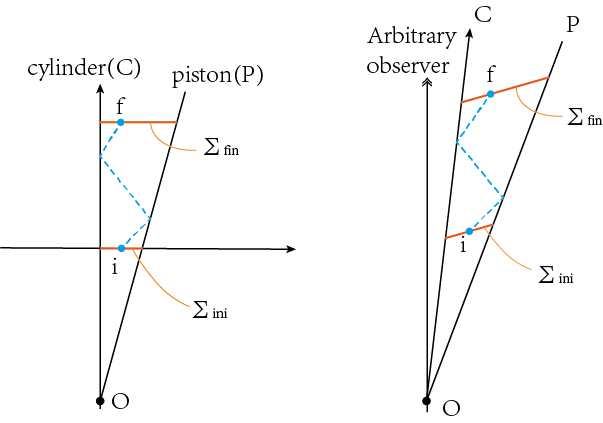}
    \captionsetup{justification=raggedright,singlelinecheck=false}
    \caption{Worldline of a particle. (Left) The dashed line denotes the worldline of a particle. Line $P$ is the worldline of the piston.  In the cylinder's frame, the worldline of the cylinder (Line $C$) coincides with the time axis ($t$-axis). An initial state $i$ lies on the $x$-axis, which is a space-like hypersurface ($\Sigma_{\mathrm{ini}}$) orthogonal to worldline $C$. The corresponding final state $f$ also lies on a space-like hypersurface ($\Sigma_{\mathrm{fin}}$, which represents the isochronous surface $t=\tau$ in the cylinder's frame). (Right) In an arbitrary inertial observer's frame, the initial states no longer belong to any isochronous surface. However, the relation that $\Sigma_{\mathrm{ini}}/\Sigma_{\mathrm{fin}}$ being orthogonal to the line $C$ is still valid.}
    \label{fig:worldline}
\end{figure}

It is worth pointing out that the initial state and the time interval $\tau$ are both defined in the cylinder's frame. 
Therefore, the initial canonical ensemble is chosen from a space-like hypersurface which is orthogonal to the worldline of the cylinder (see FIG.\ref{fig:worldline}).
Especially, when we discuss the covariance of this model, we may introduce an arbitrary inertial 
frame of reference, in which the initial state is no longer sampled from an isochronous surface (due to the relativity of simultaneity).
In this scenario, $L_i/L_f$ denotes the Minkowski's length of the initial and final hypersurfaces $\Sigma_{ini}/\Sigma_{fin}$.
These two Minkowski's lengths are Lorentz scalars, therefore we do not need to consider the Lorentz space contraction.
If we denote the velocity of this arbitrary observer as $-u$, then in this observer's frame, the velocity of the cylinder becomes $u$, and the velocity of the piston becomes $w=(u+v_p)/(1+u v_p)$.
In special relativity, the transformation of the velocity could be re-expressed with the transformation of the rapidity in a more elegant form.
In special relativity, rapidity is the inverse hyperbolic tangent of the velocity (in units of the speed of light $c$).
If we choose the natural unit that $c=1$, then we could define the rapidity of $u$ and $v_p$ as $y_u=\arctanh(u)$, $y_p=\arctanh(v_p)$.
In a Lorentz transformation, the rapidity is directly summed up.
Therefore, the velocity of the piston perceived by the observer can be re-expressed as $w=\tanh(y_p+y_u)$.

\subsection{Trajectory of a Single Particle}
In the appendix of the related work Ref.~\cite{xianghang24}, a transformation of the worldline has been introduced. 
The basic idea is to convert the polygonal worldlines into a straight auxiliary worldline, which transforms the restricted motions between boundaries into free motions \cite{moore70}.
In this article, we further discuss the physical meaning of the transformation.

In relativity, a measurement of a component of a Lorentz four-vector gives the inner product of the four-vector and a conjugate unit vector.
In this article, the metric tensor is chosen as $\eta_{\mu,\nu}=\eta^{\mu,\nu}=\mathrm{diag}\{1,-1\}$.
For the piston with four-velocity $u^\mu_p = (1-v_p^2)^{-1/2}\cdot (1,v_p)$, the covariant unit vector used to measure the momentum is a unit vector $n_{\mu,p}=(1-v_p^2)^{-1/2}(-v_p,1)$, which is orthogonal to the worldline of the piston.
The covariant unit vector to measure the energy is simply $u_{\mu,p}=\eta_{\mu \nu}u^\nu_p=(1-v_p^2)^{-1/2}(1,-v_p)$.
For a particle with four-velocity $u^\mu_{k-1}$ before the $k$-th collision with the piston, the energy and the momentum of the particle measured from the piston is
\begin{equation*}
    \begin{split}
        \mathbb{E}_{k-1}=m u^\mu_{k-1}\cdot u_{\mu,p},\\ \mathbb{P}_{k-1}=m u^\mu_{k-1} \cdot n_{\mu,p},
    \end{split}
\end{equation*}
and the four-momentum of the particle before the collision could be re-expressed as
\begin{equation*}
    P^\mu_{k-1}=m u^\mu_{k-1}=\mathbb{E}_{k-1} u^\mu_p+\mathbb{P}_{k-1} n^\mu_p.
\end{equation*}
After the collision, the energy component remains $\mathbb{E}_{k}=\mathbb{E}_{k-1}$, and the momentum component is reversed to $\mathbb{P}_{k}=-\mathbb{P}_{k-1}$, therefore the four momentum after the collision becomes 
\begin{equation*}
    P^{\mu}_{k} =\mathbb{E}_{k} u^\mu_p +\mathbb{P}_{k} n^\mu_p=m u^\mu_{k-1} =\mathbb{E}_{k-1} u^\mu_p - \mathbb{P}_{k-1} n^\mu_p.
\end{equation*}

An orthogonal reflection with respect to a hypersurface $\Sigma$ is the map that transform any Lorentz vector $\chi$ to a new vector $\chi'$ whose orthogonal component relative to $\Sigma$ is reversed, while its tangential component is preserved.
A collision with the piston serves as an orthogonal reflection with respect to the hypersurface $P$.
Since the orthogonal reflection is an involution, we may perform an extra auxiliary orthogonal reflection to preserve the four-velocity of the particle, i.e., the original four-velocity before the collision and the auxiliary four-velocity after the collision is the same.
Therefore, the original worldline segment and the auxiliary worldline segment together form a single straight line (see FIG.~\ref{fig:aux worldline}).
This framework generalizes the mirror reflection technique in the appendix of Ref.~\cite{xianghang24}, which is still valid when the reference hypersurface is not the $t$-axis.
In this framework, we no longer need to do a Lorentz boost to the piston's frame of reference.

\begin{figure*}[tb]
    \centering
    \includegraphics[width=0.8\linewidth]{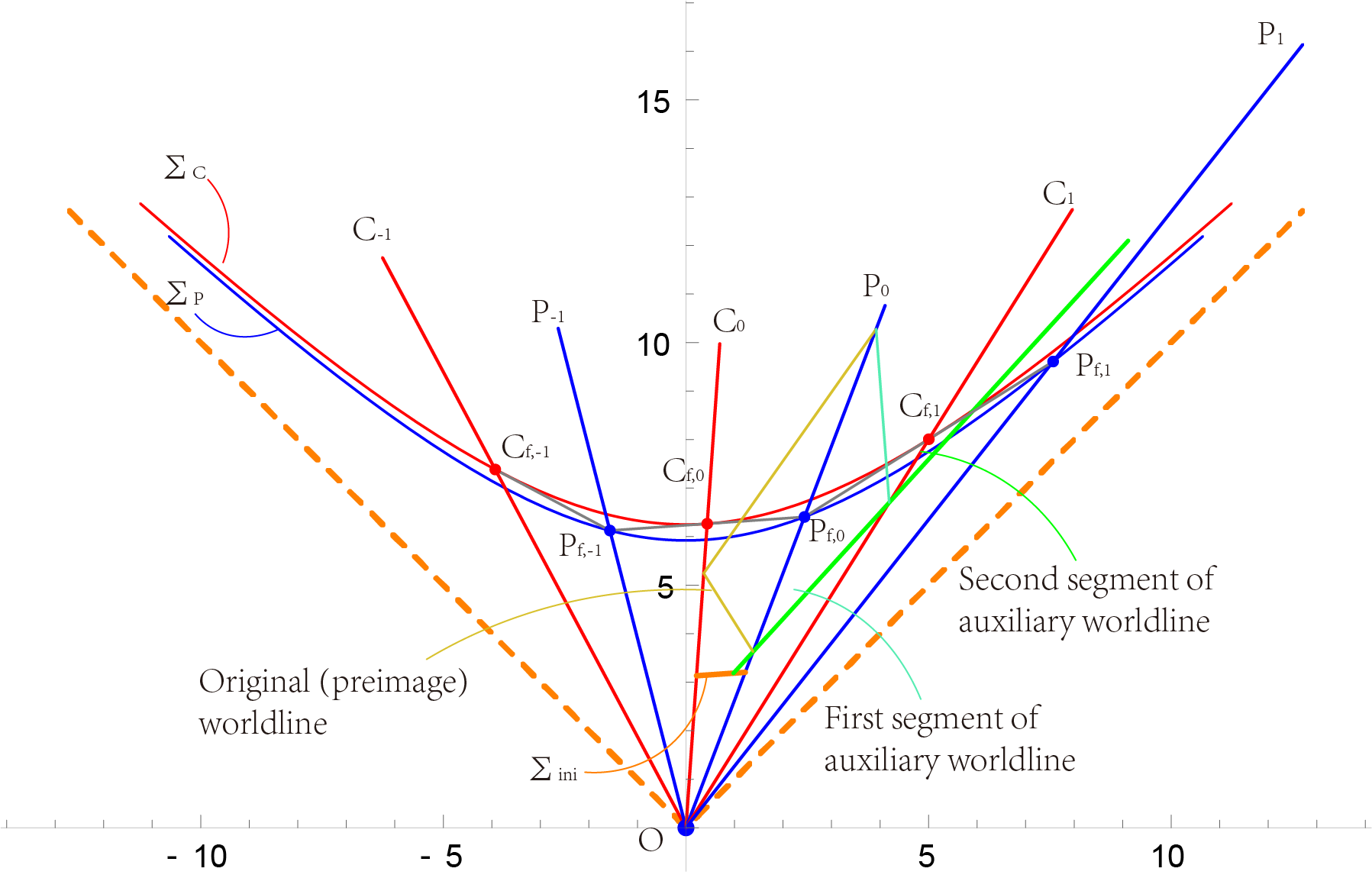}
    \captionsetup{justification=raggedright,singlelinecheck=false}
    \caption{Orthogonal reflection and auxiliary worldline. 
    This figure demonstrates the original worldline segments and the single straight auxiliary worldline after multiple orthogonal reflections.
    It also shows the results of the reflections of $C_k$'s and $P_k$'s.
    Meanwhile, the hyperbolas $\Sigma_{\mathrm{C}}$, $\Sigma_{\mathrm{P}}$ denote the events that shares the same space-time interval relative to $O$.
    The intersection between $C_k$ and $\Sigma_{\mathrm{C}}$ is $C_{\mathrm{f,k}}$, and the intersection between $P_k$ and $\Sigma_{\mathrm{P}}$ is $P_{\mathrm{f,k}}$, which are denoted by the red and blue dots.
    In this figure, the velocity $-u$ of the observer is nonzero, therefore, $C_0$ does not coincide with the $t$-axis.
    For the same reason, the $\Sigma_{\mathrm{ini}}$ is no longer isochronous.}
    \label{fig:aux worldline}
\end{figure*}

An orthogonal reflection is an isometry.
For a pair of events $e_0,e_1$ and their image $e'_0,e'_1$, the space-time interval $|e_0-e_1|^2$ of the separation is unchanged during the reflection.

During multiple collisions, we may introduce multiple orthogonal reflections with respect to either the (auxiliary) hypersurface $C_k$ (denoting the $k$-th collision with the cylinder) or the (auxiliary) hypersurface $P_k$ (denoting the $(k+1)$-th collision with the piston).
For example, after the first collision with the piston, we introduce the first auxiliary worldline of $C$ as $C_1$.
Then, to deal with the second collision between the auxiliary worldline of the particle and $C_1$, we perform the second orthogonal reflection with respect to $C_1$, which maps $P$ to the second auxiliary worldline $P_1$.
If the third collision with $P_1$ happens, the third reflection with respect to $P_1$ will similarly reflect $C_1$ to $C_2$.
If we denote the original worldline as $C_0$ and $P_0$ (for unification), then the reflection with respect to $C_k$ maps $P_{k-1}$ to $P_{k}$, and the reflection with respect to $P_k$ maps $C_k$ to $C_{k+1}$.
If the particle collides with the cylinder first, we reflect $P_0$ with respect to $C_0$.
If we follow the same rule to label the auxiliary worldline of $C$ and $P$, we may label the reflection of $P_0$ with respect to $C_0$ as $P_{-1}$.
Similarly, we may reflect $C_0$ with respect to $P_{-1}$ to obtain $C_{-1}$, reflect $P_{-1}$ with respect to $C_{-1}$ to obtain $P_{-2}$, and so on (see FIG.~\ref{fig:aux worldline}). 

For any preimage $e_0$ event, the image after the $k$-th reflection is denoted as $e_k$ (if the first reflection is a reflection with respect to $C_0$, and the total number of reflections is $|k|$, we convent the order of the image is $k=-|k|$).

The intersection between $C$ and $P$ is the fixed point during the reflections, which is denoted as $O$.
$O$ lies on every reflection surface, therefore the space-time separation between $O$ and a preimage $e_0$ equals the space-time separation between $O$ and an arbitrary $k$-th order image $e_{k}$.
Especially, if we choose the preimage events as the intersections of $\Sigma_{\mathrm{fin}}$ and the worldline $C_0$ and $P_0$, and denote them as $C_{\mathrm{f,0}}/P_{\mathrm{f,0}}$, and their $k$-th order images as $C_{\mathrm{f,k}}/P_{\mathrm{f,k}}$, then 
we will find that
\begin{equation*}
    \begin{split}
        OC_{\mathrm{f,0}}\cdot OC_{\mathrm{f,0}} &=OC_{\mathrm{f,k}}\cdot OC_{\mathrm{f,k}},\\
        OP_{\mathrm{f,0}}\cdot OP_{\mathrm{f,0}} 
        &=OP_{\mathrm{f,k}}\cdot OP_{\mathrm{f,k}}.
    \end{split}
\end{equation*}
As a result, every $C_{\mathrm{f,k}}/P_{\mathrm{f,k}}$ lies on a hyperbolic surface centered at $O$.
We denote the hypersurfaces as $\Sigma_{\mathrm{C}}$ and $\Sigma_{\mathrm{P}}$,
and the Minkowski distances between $\Sigma_{{\mathrm{C}}}/\Sigma_{{\mathrm{P}}}$ and $O$ are
\begin{equation}\label{eqn:distance}
    \begin{split}
        S_C=|OC|&=\tau+\frac{L_i}{v_p}, \\
        S_P=|OP|&=\sqrt{1-v_p^2}(\tau+\frac{L_i}{v_p}).
    \end{split}
\end{equation}
This treatment is similar to the hyperbolic coordinate techniques introduced in Ref.~\cite{koehn12}.

The slope of the auxiliary worldline could be determined as follows: 
an orthogonal reflection keeps the absolute value of the relative rapidity.
If we denote the rapidity of the piston to be $y_p=\arctanh(v_p)$, then, for the first collision with the piston, the corresponding rapidity of the auxiliary worldline $C_{1}$ is $2 y_p$.
Similarly, the rapidity of the auxiliary $P_{1}$ is $3y_p$.
By performing the transformation multiple times, we conclude that the rapidity of $C_k$ is $2k y_p$, and the rapidity of $P_k$ is $(2k+1)y_p$.

The image $C_{\mathrm{f,k}}$ is the intersection between $\Sigma_C$ and $C_k$, while the image $P_{\mathrm{f,k}}$ is the intersection between $\Sigma_P$ and $P_k$.
Therefore, for the $k$-th auxiliary $C_{\mathrm{f,k}}/P_{\mathrm{f,k}}$, the temporal coordinate $t(C_{\mathrm{f,k}})/t(P_{\mathrm{f,k}})$ and spatial coordinate $x(C_{\mathrm{f,k}})/x(P_{\mathrm{f,k}})$ are
\begin{equation}\label{eqn:coordinate rest frame}
    \begin{split}
        [t(C_{\mathrm{f,k}}),x(C_{\mathrm{f,k}})]&=S_C [\cosh(2k y_p), \sinh(2k y_p)],\\
        [t(P_{\mathrm{f,k}}),x(P_{\mathrm{f,k}})]&=S_P [\cosh((2k+1) y_p), \sinh((2k+1) y_p)].
    \end{split}
\end{equation}
The coordinates of the auxiliary $C_{\mathrm{f,k}}/P_{\mathrm{f,k}}$ could be used to determine the total number of collisions.

For an observer moving with velocity $-u$ relative to the cylinder, we may simply perform a Lorentz transformation centered at $O$ with the rapidity $y_u=\mathrm{arctanh}(u)$, and the influence is to boost the rapidity of every (auxiliary) worldline with the rapidity $n y_p$ to a new (auxiliary) worldline with the rapidity $y_u +n y_p$ (see FIG.~\ref{fig:aux worldline}).
The space-time interval is invariant under a Lorentz transformation, therefore $S_{C}/S_{P}$ are unchanged.
The intersections between the auxiliary worldline and $\Sigma_{\mathrm{C}}/\Sigma_{\mathrm{P}}$ are thus boosted to
\begin{equation}
    \begin{split}
        &[t(C_{\mathrm{f,k}}),x(C_{\mathrm{f,k}})]\\
        =&S_C [\cosh(2k y_p+y_u), \sinh(2k y_p+y_u)],\\
        &[t(P_{\mathrm{f,k}}),x(P_{\mathrm{f,k}})]\\
        =&S_P [\cosh((2k+1) y_p+y_u), \sinh((2k+1) y_p+y_u)].
    \end{split}
\end{equation}
These points determines the boundary of (auxiliary) $\Sigma_{\mathrm{fin}}$, and correspondingly, the $\Sigma_{\mathrm{ini}}$ is also transformed to another space-like hypersurface.
Denote the intersection between $\Sigma_{\mathrm{ini}}$ and $C$ (and $P$) as $C_i:L_i/v_p[1,0]$ (and $P_i:\sqrt{1-v_p^2}L_i/v_p[\cosh(y_p), \sinh(y_p)]$), then the transformed boundary becomes $C_i:L_i/v_p[ \cosh(y_u), \sinh(y_u)]$ and $P_i:\sqrt{1-v_p^2}L_i/v_p[\cosh(y_p+y_u), \sinh(y_p+y_u)]$ (see FIG.~\ref{fig:aux worldline}).

\subsection{\label{subsec:verification} Classification of the Number of Collisions in the State Space}
For a particle that collides with the piston first, if the initial velocity is $v_i=\tanh(y_i)$, here $y_i$ is the initial rapidity, then the rapidity after the $n$-th collision is
\begin{equation}
    y_n=\begin{cases}
        y_i-2 k y_p,& n=2k\\
        -y_i +2k y_p + 2 y_u, & n=2k-1.
    \end{cases}
\end{equation}
If a particle collide with the cylinder first, the rapidity after the first collision is $-y_i+2 y_u$, which satisfies the expression of $y_n\bigm|_{n=-1}$.
Therefore, we adopt the convention that if the particle collides with the cylinder first, we take the negative number of the collision time as the effective $n$ in the expression of final rapidity.

Once we know the initial state (position and velocity) and the number of collisions, we can deduce the final state of the particle.
The key question in deriving the dynamics of this collision model is to determine the number of collisions for a given protocol and initial state.
This classification can also be done geometrically with the auxiliary worldline.

\begin{figure*}[tb]
    \centering
    \includegraphics[width=0.8\linewidth]{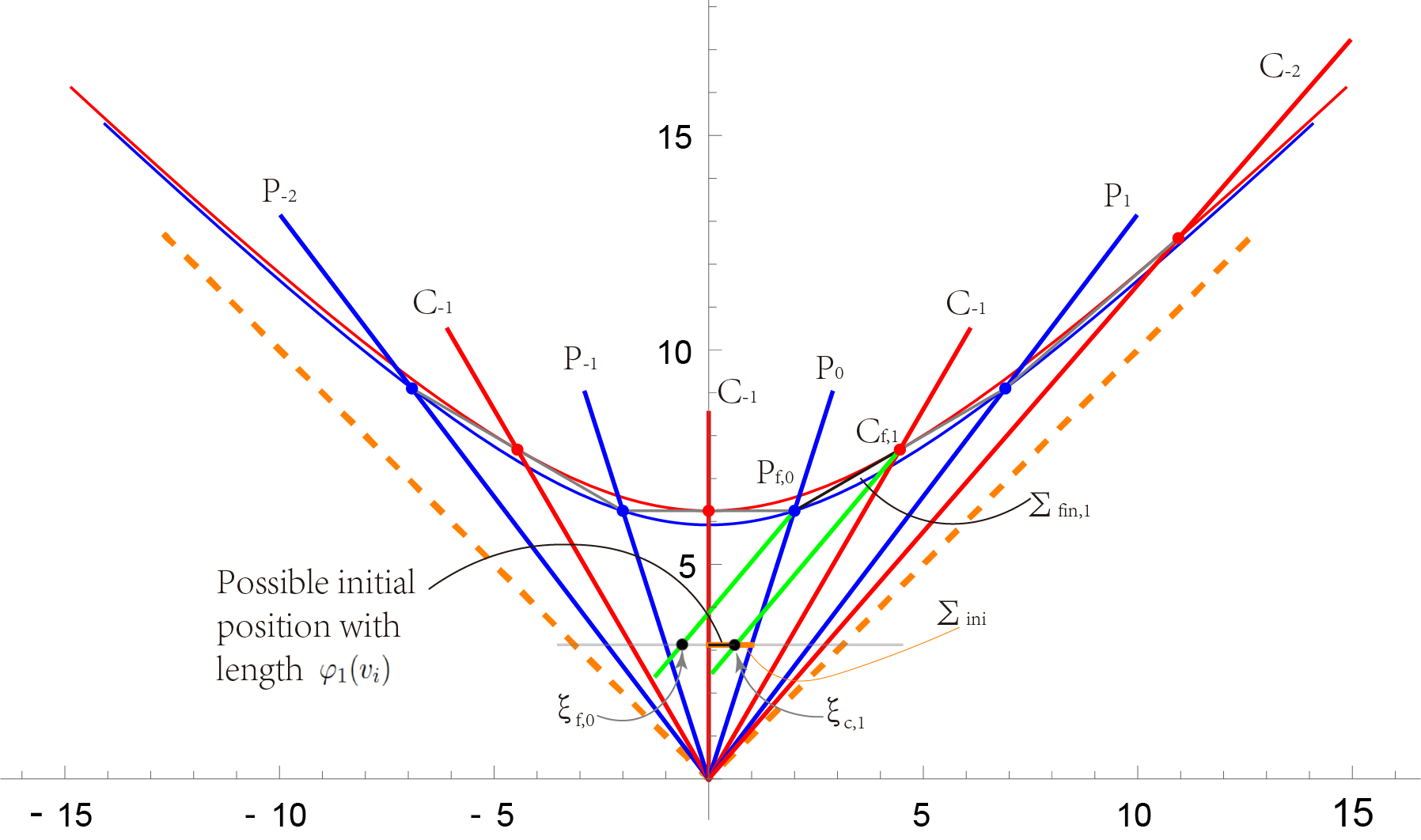}
    \captionsetup{justification=raggedright,singlelinecheck=false}
    \caption{The ``shadow" of $\Sigma_{\mathrm{fin,n}}$ on $\Sigma_{\mathrm{ini}}$ is the possible initial position for a particle with initial velocity $v_i$ to collide $n$ times.
    The ``parallel light" is represented by the green lines, and the slope of the ``parallel light" is determined by $v_i$.
    In this example, the boundaries of the first ``shadow" is $\xi_{\mathrm{P,0}}(v_i)$ and $\xi_{\mathrm{C,1}}(v_i)$, which depend on the velocity of the ``parallel light".
    They determine the spatial boundaries of domain $D_n$ in the state space.
    The Minkowski's length of the ``shadow" projected on $\Sigma_{\mathrm{ini}}$ is a Lorentz scalar, therefore we may calculate this value in the laboratory reference frame ($u=0$).}
    \label{fig:shadow and overlap}
\end{figure*}

\begin{figure}[ht]
    \centering
    \includegraphics[width=0.9\linewidth]{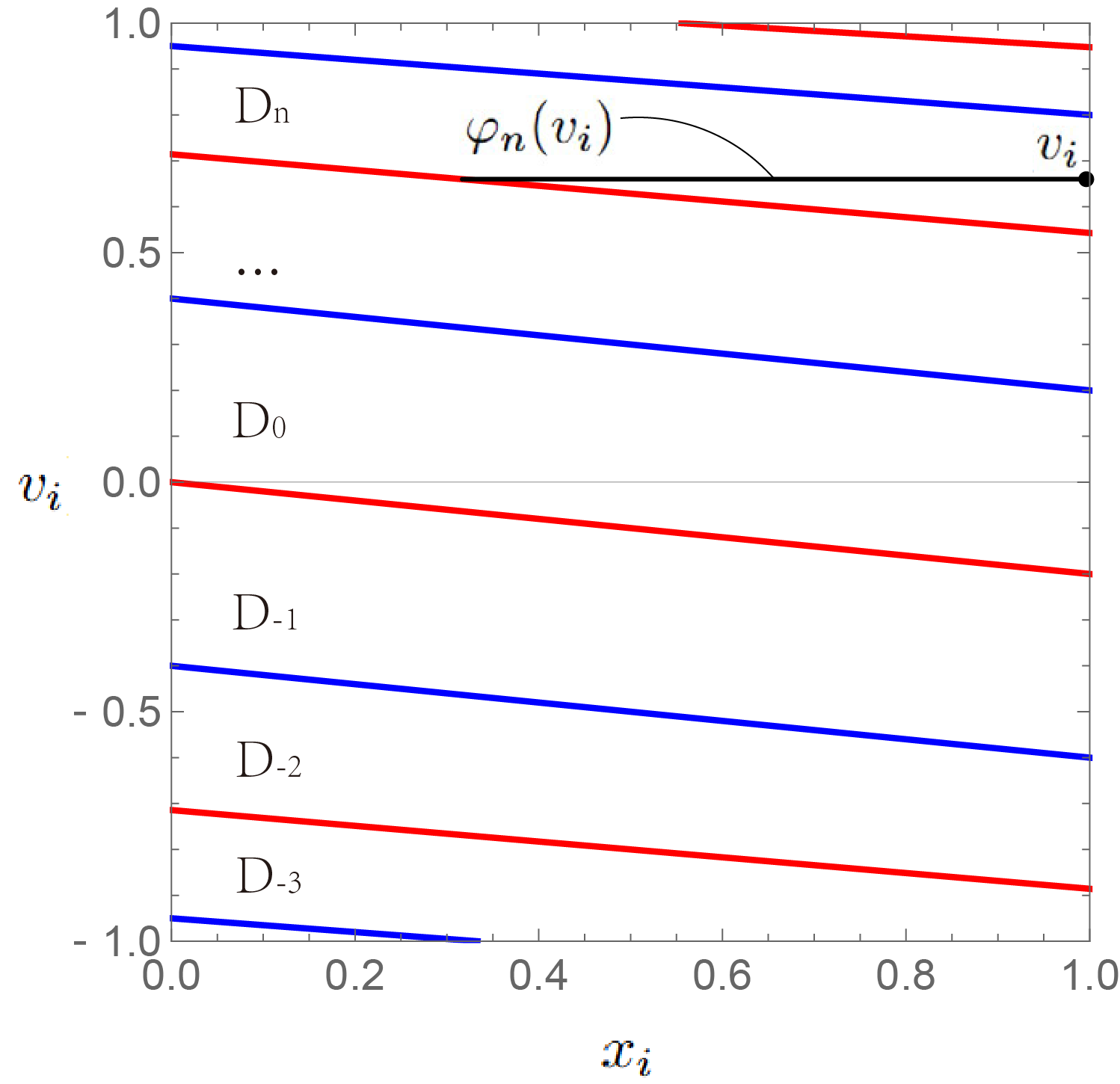}
    \captionsetup{justification=raggedright,singlelinecheck=false}
    \caption{Separation of domain $D_n$ and the physical meaning of the overlap function $\varphi_n(v_i)$.}
    \label{fig:domain and overlap}
\end{figure}

In case that the particle happen to collide exactly $n$ times with the boundaries, the (auxiliary) final state must lie on the segment $C_{\mathrm{f,k}}-P_{\mathrm{f,k}}$ if $n=2k$, or on the segment $P_{\mathrm{f,k-1}}-C_{\mathrm{f,k}}$ if $n=2k-1$.
We denote the final (auxiliary) space-like hypersurface as $\Sigma_{\mathrm{fin,n}}$.
If we hope to classify the state space ($x-v$ space) into regions of different collision numbers, then for a given initial velocity $v_i$, we must figure out what is the range of the initial position allowed for the particle that ensures the particle stops on $\Sigma_{\mathrm{fin,n}}$.
The procedure is simple: 
We project the (auxiliary) final hypersurface $\Sigma_{\mathrm{fin,n}}$ along the initial velocity $v_i$ onto the initial hypersurface $\Sigma_{\mathrm{ini}}$.
The region projected on $\Sigma_{\mathrm{ini}}$ is the allowed initial position for the particle to end up on $\Sigma_{\mathrm{fin,n}}$.
This projection is like the ``shadow" of $\Sigma_{\mathrm{fin,n}}$ under a ``parallel light" whose direction is determined by the initial velocity $v_i$ (see FIG.~\ref{fig:shadow and overlap}).
The left and right boundaries of the ``shadow" separate the initial state space into regions $D_n$ with different numbers of collisions $n$ (see FIG.~\ref{fig:domain and overlap}).
When we hope to calculate the distribution function, we can do integration on different state space regions separately.

The overlap between the ``shadow" and $\Sigma_{\mathrm{ini}}$ is a time-like hypersurface.
In this (1+1) dimensional model, the hypersurface is simply a line segment (in FIG.~\ref{fig:shadow and overlap}, the black segment, which is the set of all possible initial positions, is the overlap here).
We denote the Minkowski's length of the space-like line segment as the overlap function $\varphi_{n}(v_i)$ \cite{xianghang24} for a given initial velocity $v_i$.
The expression of the overlap function is calculated (see Appendix~\ref{appendix:overlap}).
We would like to emphasize that the solution to the motion of the particle is encoded in this overlap function.

With the expression of the boundaries of the ``shadow", i.e., Eq.~(\ref{eqn:shadow boundaries}), we may separate the state space $(x_i,v_i)$ into different regions $D_n$ with different numbers of collisions $n$, see FIG.~\ref{fig:domain and overlap}.
For a given initial velocity and position, we could locate the position of the state in $D_n$ and find the number of collisions.
The dynamics of the piston model is completely solved by this geometrical method.
The overlap function shows the fraction of the initial position that collide exactly $n$ times for a given initial velocity $v_i$, which is helpful in calculating the ensemble average.

\section{\label{sec:work}Joint Distribution of Four-Vector work in the relativistic piston model}

In special relativity, the four-vector work is defined as the change of four-momentum caused by the driving protocol.
In this section, we focus on the four-vector work on the trajectory level, and calculate the joint distribution of four-vector work perceived by an arbitrary inertial observer.
\subsection{\label{subsec:plw}Lorentz Covariant Work Distribution}
In this relativistic case, the initial state satisfies Maxwell-J\"uttner distribution at the inverse temperature $\beta$ \cite{jut11}\\
\begin{equation}
    \begin{split}
        J\ddot{u}_p(x',p')=\frac{1}{2K_1(\beta mc^2)L_i}e^{-\beta{mc^2}\sqrt{1+(\frac{p'}{mc})^2}},
    \end{split}
\end{equation}
where $K_1$ is the modified Bessel function of the second kind, $(x',p')$ the initial position and momentum of the given particle observed from the cylinder frame, $L_i$ the initial length of the cylinder. This distribution is equivalent to the following distribution expressed in the position-rapidity space.\\
\begin{equation}
    \begin{split}
        f(x',y'_i)=\frac{1}{2K_1(\beta)L_i}\cosh(y'_i)e^{-\beta \cosh(y'_i)}.
    \end{split}
\end{equation}

For simplicity, we set $L_i$ to be $1$. Lorentz transformation gives the relation $y'_i=y_i-y_u$, and thus, the four-work distribution can be calculated as
\begin{equation}
    \begin{split}
    F_L(W^0,W^1)=\int_{0}^{L_i}\mathrm{d}x'\int_{-\infty}^{\infty}\mathrm{d}y_i\frac{\exp\left(-\beta \cosh(y_i-y_u)\right)}{2K_1(\beta)}\times\\ \cosh(y_i-y_u)\delta\left(W^0 - W^0_{\tau}(x, y_i)\right)\delta\left(W^1 - W^1_{\tau}(x, y_i)\right),
    \end{split}
\end{equation}
where the subscript $L$ in $F_L(W^0,W^1)$ stands for the Lorentz covariance.

The four-vector work done by the particle that has collided n times with both the piston and the cylinder wall in the given time period $\tau$ is
 \begin{equation}\label{eqn:wzeroy}
        W^0_\tau = 
        \begin{cases}
            2\sinh(y_i-ky_p)\sinh(ky_p), & n=2k \\
            2\sinh(ky_p+y_u)\sinh(y_i-y_u-ky_p), & n=2k-1
        \end{cases}
\end{equation}
 \begin{equation}
        W^1_\tau = 
        \begin{cases}
            2\cosh(y_i-ky_p)\sinh(ky_p), & n=2k \\
            2\cosh(ky_p+y_u)\sinh(y_i-y_u-ky_p), & n=2k-1.
        \end{cases}
\end{equation}
The above expressions indicate that the zeroth component of the four-vector work is the energy change in classical cases, and the rest three components stand for the classical momentum change.

Utilizing the results obtained in the previous discussion, the integration can be separated into different parts according to the collision count, analytically expressed as
\begin{equation}
    \begin{split}
    F_L(W^0,W^1)=\sum_n\int_{D_n}\mathrm{d}x'\mathrm{d}y_i\frac{\exp\left(-\beta \cosh(y_i-y_u)\right)}{2K_1(\beta)}\times \\
    \cosh(y_i-y_u)\delta\left(W^0 - W^0_{\tau}(x, y_i)\right)\delta\left(W^1 - W^1_{\tau}(x, y_i)\right).
    \end{split}
\end{equation}

Since each part of the integrand is independent of $x'$, by integrating $x'$ out first and combine the result with the overlap function derived in the previous section, each component of the summation yields
\begin{equation}\label{eqn:L}
    \begin{split}
    \int_{D_n}\mathrm{d}y_i\frac{\exp\left(-\beta \cosh(y_i-y_u)\right)\cosh(y_i-y_u)}{2K_1(\beta)}\times \\ \delta\left(W^0 - W^0_{\tau}(x, y_i)\right)\delta\left(W^1 - W^1_{\tau}(y_i)\right)\varphi_n\left(\tanh(y_i-y_u)\right),
    \end{split}
\end{equation}
where $\varphi_n\left(\tanh(y_i)\right)$ is the overlap function in Eq. (\ref{eqn:overlap}). 

\begin{figure}[ht]
    \centering
    \includegraphics[width=0.95\linewidth]{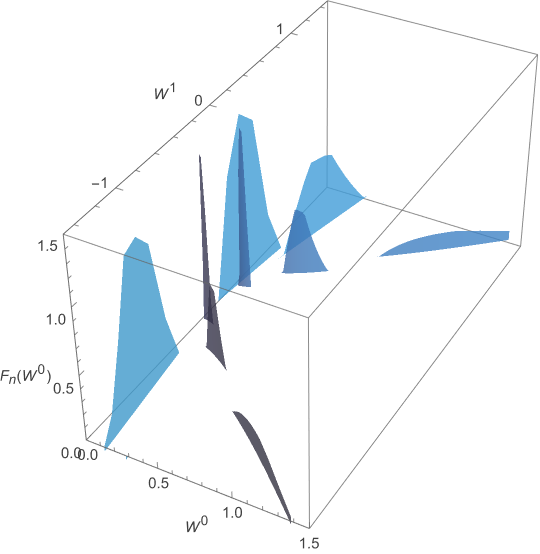}
    \captionsetup{justification=raggedright,singlelinecheck=false}
    \caption{Four-vector work distribution $P_L(W^0,W^1)$. 
    Here, the parameters are chosen to be $L_i=1,v_p=0.1,\tau=10,\beta=0.8,u=0$.
    The height of each sheet denotes the magnitude of the $\delta$ function. The work distribution is concentrated on some line segments in the $(W^0, W^1)$ space. 
    The relation between $P_L(W^0,W^1)$ and $P_n(W^0)$ is given in Eq.~(\ref{eqn:P Lorentz}).}
    \label{fig:lorentz}
\end{figure}

The remaining calculation involves a Dirac $\delta$ function. See Appendix \ref{appendix:er} for exact analytical derivations and results. A more intuitive representation of the joint work distribution is illustrated in FIG.~\ref{fig:lorentz}.

\begin{figure}[htbp]
    \centering
    \begin{subfigure}[b]{0.22\textwidth}
        \centering
        \includegraphics[width=\textwidth]{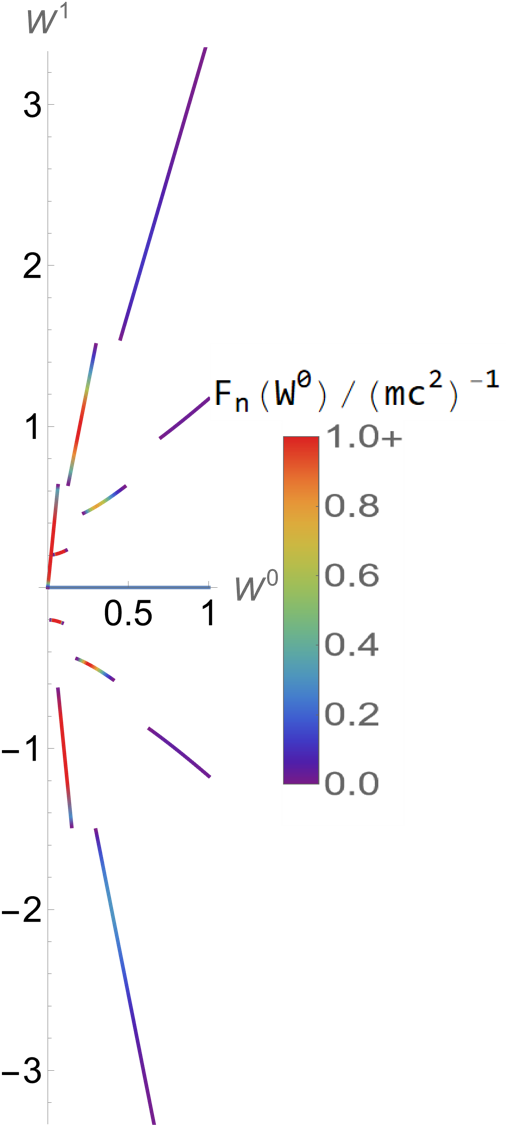}
        \caption{u=0}
        \label{fig:u=0}
    \end{subfigure}
    \begin{subfigure}[b]{0.235\textwidth}
        \centering
        \includegraphics[width=\textwidth]{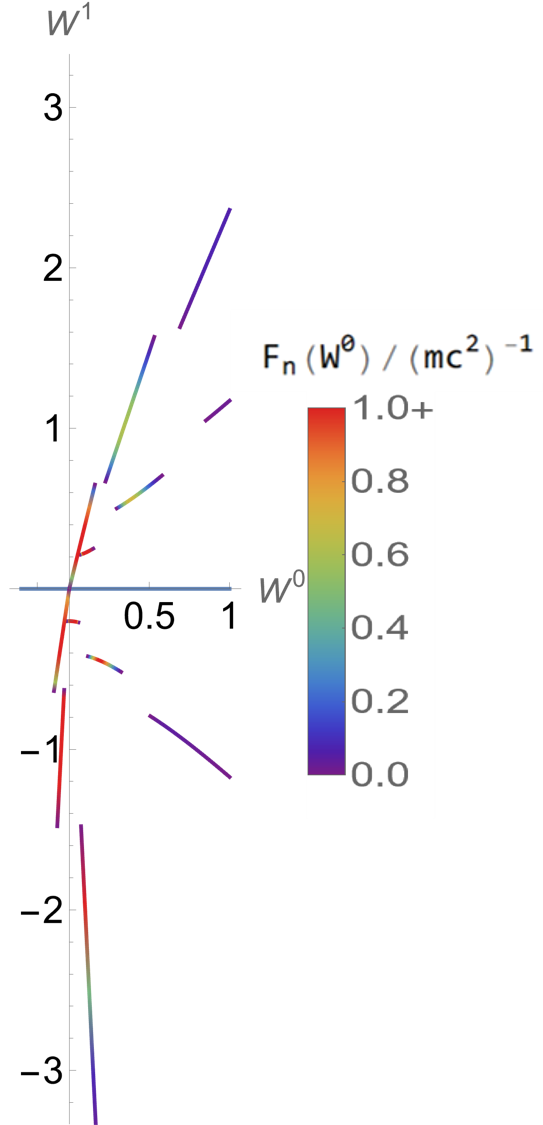}
        \caption{u=0.15c}
        \label{fig:u=0.15c}
    \end{subfigure}
    \captionsetup{justification=raggedright,singlelinecheck=false}
    \caption{Joint distribution in different frames of reference.
    Here, the parameters are chosen to be $L_i=1,v_p=0.1,\tau=10,\beta=5$.
    Observers are moving with velocities $-u$ relative to the cylinder. The color of the line denotes $P_n(W^0)$, and the position of the line is determined by $\delta(W^1-W^1_\tau (W^0))$ in the $(W^0,W^1)$ space.
     The relation between $P_L(W^0,W^1)$ and $P_n(W^0)$ is given in Eq.~(\ref{eqn:P Lorentz}).}
    \label{fig:LT of work distribution}
\end{figure}

For observers with different velocity $u$, the distribution of work may vary.
However, there are some features unchanged for both even and odd numbers of collisions (see FIG.~\ref{fig:LT of work distribution}).
For even $n$'s, the four-vector work distribution remains lying on the same hyperbolas with the constant spacelike distance to the origin.
The positions of the non-zero parts on the hyperbolas change with the velocity of the observer.
For odd $n$'s, the linear relation between $W^0$ and $W^1$ remains, only with the slope changed due to the Lorentz transformation.
The joint distribution of four-vector work displays some signatures of the Lorentz covariance: the Lorentz transformation is a linear transformation maintaining the spacetime interval.

\subsection{\label{subsec:plg}Galileo Covariant Work Distribution}
Consider a piston model with exactly the same setup, except that the particles are no longer relativistic but obey the Newtonian mechanics. 

The dynamics of the system can be resolved by repeating the `worldline' method introduced in the previous section. Given the initial velocity $v$, the velocity after colliding $n$ times can be directly derived:
\begin{equation}
    v_n=
    \begin{cases}
    v_i+2(k-1)u-2kw, n=2k-1~(k\in \mathbb{Z})\\
    v_i+2ku-2kw, n=2k~(k\in \mathbb{Z}).
    \end{cases}
\end{equation}

The intersection between the worldlines of the piston and the cylinder wall is $(-{Lu}(w-u)^{-1},-{L}({w-u})^{-1})$. For a particle that collides exactly n times with both the piston and the cylinder wall within time $\tau$, the inverse slope range of the worldlines is $(\tanh(y_u+ny_p),\tanh(y_u+(n+1)y_p))$. Thus, the range of the initial velocity can be expressed as
\begin{widetext}
    \begin{equation}
        \begin{split}
        v_i\in\left(\left[\left(\tau+\frac{L_i}{w-u}\right)(u+nv_p)-\frac{L_iu}{w-u}-x\right]\cdot\frac{1}{\tau},\left[\left(\tau+\frac{L_i}{w-u}\right)\left(u+(n+1)v_p\right)-\frac{L_iu}{w-u}-x\right]\cdot\frac{1}{\tau}\right).
        \end{split}
    \end{equation}
\end{widetext}
The domain is thus separated into parallelogram regions and the overlap function reads\\
\begin{widetext}
    \begin{equation}\label{eqn:goverlap}
        \begin{split}
        \varphi_n(v_i)=
            \begin{cases}
           -\left(\tau+\frac{L_i}{w-u}\right)(u+(n-1)v_p)+\frac{L_iw}{w-u}+v_i\tau,\\
           \qquad\qquad v_i\in\left[\left[\left(\tau+\frac{L_i}{w-u}\right)(u+nv_p)-\frac{L_iw}{w-u}\right]\cdot\frac{1}{\tau},\left[\left(\tau+\frac{L_i}{w-u}\right)(u+nv_p)-\frac{L_iu}{w-u}\right]\cdot\frac{1}{\tau}\right]\\
           L_i,\\\qquad\qquad v_i\in\left[\left[\left(\tau+\frac{L_i}{w-u}\right)(u+nv_p)-\frac{L_iu}{w-u}\right]\cdot\frac{1}{\tau},\left[\left(\tau+\frac{L_i}{w-u}\right)\left(u+(n+1)v_p\right)-\frac{L_iw}{w-u}\right]\cdot\frac{1}{\tau}\right]\\
           \left(\tau+\frac{L_i}{w-u}\right)(u+nv_p)-\frac{L_iu}{w-u}-v_i\tau, \\
           \qquad\qquad v_i\in\left[\left[\left(\tau+\frac{L_i}{w-u}\right)\left(u+(n+1)v_p\right)-\frac{L_iw}{w-u}\right]\cdot\frac{1}{\tau},\left[\left(\tau+\frac{L_i}{w-u}\right)\left(u+(n+1)v_p\right)-\frac{L_iu}{w-u}\right]\cdot\frac{1}{\tau}\right].
            \end{cases}
        \end{split}
    \end{equation}
\end{widetext}
In comparison with the relativistic case (Eq.~\ref{eqn:overlap}), this non-relativistic result is much simpler.

As the particles obey Maxwellian distribution, the work distribution reads
\begin{equation}
    \begin{split}
    F_G(W^0,W^1)=\sum^n\int\mathrm{d}x_i\mathrm{d}v_i\frac{\sqrt{\beta}\exp\left(-\frac{1}{2}(v_i-u)^2\right)}{\sqrt{2\pi}L_i}\times\\\delta(W^0-W^0_{\tau})\delta(W^1-W^1_{\tau}),
    \end{split}
\end{equation}
where the subscript $G$ in $F_G(W^0,W^1)$ stands for Galileo covariance.

Denote $F_{n}(W^0)$ as
\begin{equation}
    \begin{split}
        \int\mathrm{d}x_i\mathrm{d}v_i\frac{\sqrt{\beta}\exp\left(-\frac{1}{2}(v_i-u)^2\right)}{\sqrt{2\pi}L_i}\delta(W^0-W^0_{\tau}),
    \end{split}
\end{equation}
and thus $F_G$ can be expressed as
\begin{equation}\label{eqn:P Galileo}
    \begin{split}
        F_G(W^0,W^1)=\sum^n F_n(W^0)\delta(W^1-W^1_{\tau}).
    \end{split}
\end{equation}

By treating odd $n$ and even $n$ separately, this distribution can be expressed analytically. For simplicity, set $L_i$ to be $1$.

For n as an odd integer, let $n=2k-1$. As the integrand is independent of $x$, by integrating $x$ out first, $F_{2k-1}$ can be expressed as
\begin{equation}\label{eqn:gpodd}
    \begin{split}
        F_{2k-1}(v_i)=\frac{\sqrt{\beta}}{\sqrt{2\pi}L}\varphi_{2k-1}(v_i)\frac{\exp\left(-\frac{1}{2}(v_i-u)^2\right)}{2kw-2(k-1)u},
    \end{split}
\end{equation}
and
\begin{equation}\label{eqn:gdodd}
    \begin{split}
        W^1_{\tau}=2v_i-2kw+2(k-1)u.
    \end{split}
\end{equation}
Notice that $v_i$ is a function of $W^0$:
\begin{equation}\label{gvodd}
    \begin{split}
        v_i=\frac{W^0+2\left((k-1)u-kw\right)^2}{2\left(kw-(k-1)u\right)},
    \end{split}
\end{equation}
thus $F_{2k-1}$ can be derived. The exact result is given in Appendix \ref{appendix:g}, see Eq. (\ref{eqn:luapw1}) and Eq. (\ref{eqn:2odddelta}).

For $n$ as an even integer, let $n=2k$. The calculation process is similar to the case where $n$ is an odd number, except that the change of momentum is $2kw-2ku$ due to the fact that $v_{2n}$ points at the same direction as $v_i$. The exact work distribution for colliding $2k$ times with both the cylinder wall and the piston can be derived, and the exact distribution is given in Appendix \ref{appendix:g} (see Eq. (\ref{eqn:luapw2}) and Eq. (\ref{eqn:2evendelta})).

\begin{figure}
    \centering
    \includegraphics[width=0.95\linewidth]{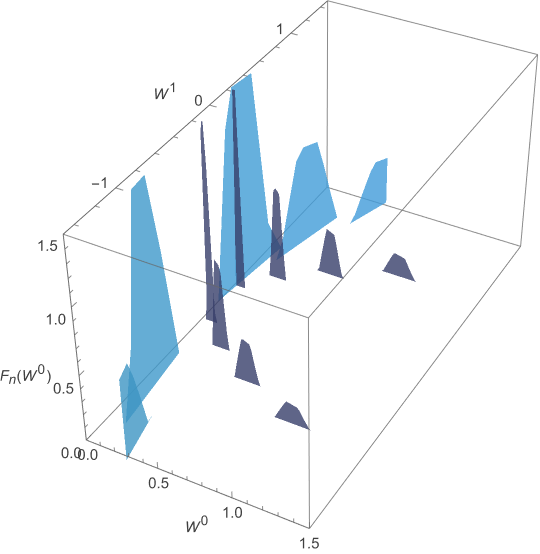}
    \captionsetup{justification=raggedright,singlelinecheck=false}
    \caption{Four-vector work distribution for non-relativistic particle. 
    Here, the parameters are chosen to be $L_i=1,v_p=0.1,\tau=10,\beta=0.8,u=0$.
    The height of each sheet denotes the magnitude of the $\delta$ function. The work distribution is also concentrated on some line segments in the $(W^0,W^1)$ space. 
    The relation between $F_L(W^0,W^1)$ and $F_n(W^0)$ is given in Eq.~(\ref{eqn:P Galileo}).
    We can see that on the sheets representing odd numbers of collisions, the $W^1$ components are some constants, which significantly differs from the special relativistic results.}
    \label{fig:galileo}
\end{figure}

A more intuitive representation of the joint work distribution is illustrated in FIG.~\ref{fig:galileo}.

\subsection{\label{subsec:landg}Consistency and Differences between Lorentz and Galileo Covariant Work Distributions}
For a relativistic system, when taking the low speed limit, the results attained through relativistic mechanics are expected to coincide with Newtonian mechanics. This remains valid in the relativistic piston model. 
The Lorentz covariant four-vector work distribution derived in the previous paragraphs is supposed to approach that of the Galileo covariant four-vector work distribution. In the following, we will check this.

To start with, consider the separation of domains. By taking the low speed limit, $\sinh(ny_p)$ in Eq. (\ref{eqn:overlap}) becomes $nv_p$, and $\cosh(ny_p)$ becomes $1$. Obviously, the separation of domain in Eq. (\ref{eqn:overlap}) is equivalent to the separation in Eq. (\ref{eqn:goverlap}). As both overlap functions form trapezoidal regions on the $x-v$ plane, this equivalence indicates that the overlap function of the Lorentz covariant case is consistent with the overlap function in the Galileo covariant case.

Next, examine each component of the distribution. Take $F_{2k-1}$ and $F_{2k}$ in the Lorentz covariant distribution. If each counterpart in the Lorentz and Galileo covariant distribution is equivalent to each other, then the consistency between the two covariant models can be verified. This consistency holds true analytically, see Appendix \ref{appendix:d} for detailed demonstration.

\begin{figure}[htbp]
    \centering
    \begin{subfigure}[b]{0.41\textwidth}
        \centering
        \includegraphics[width=\textwidth]{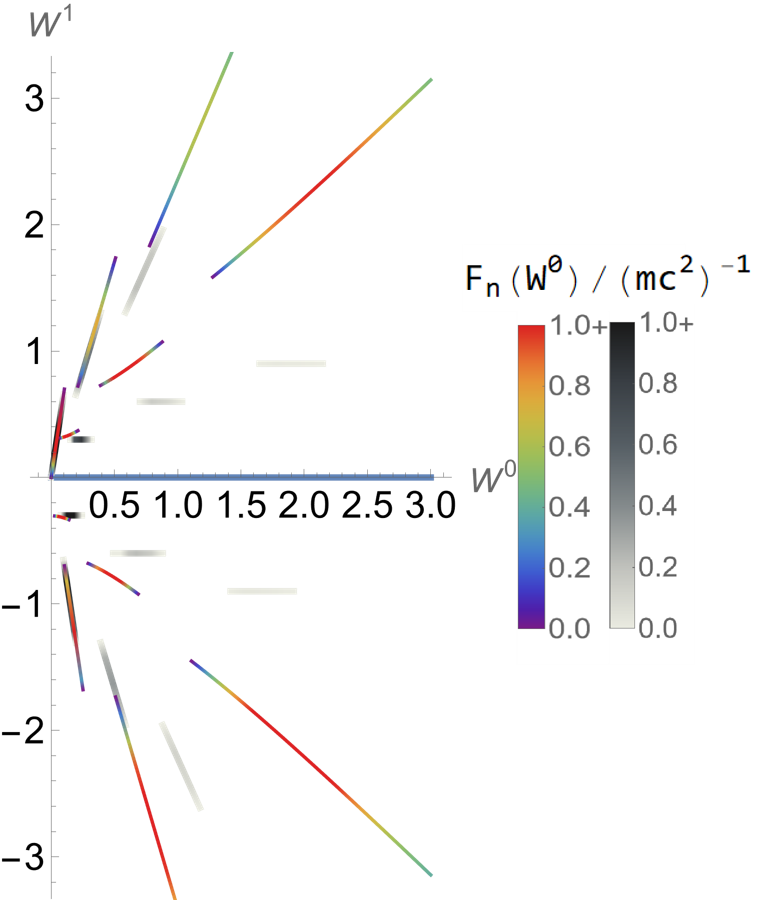}
        \caption{$Z_f/Z_i=2.7,v_p=0.15c,\beta=1.2,u=0$.}
        \label{fig:compare high T}
    \end{subfigure}
    \\
    \begin{subfigure}[b]{0.41\textwidth}
        \centering
        \includegraphics[width=\textwidth]{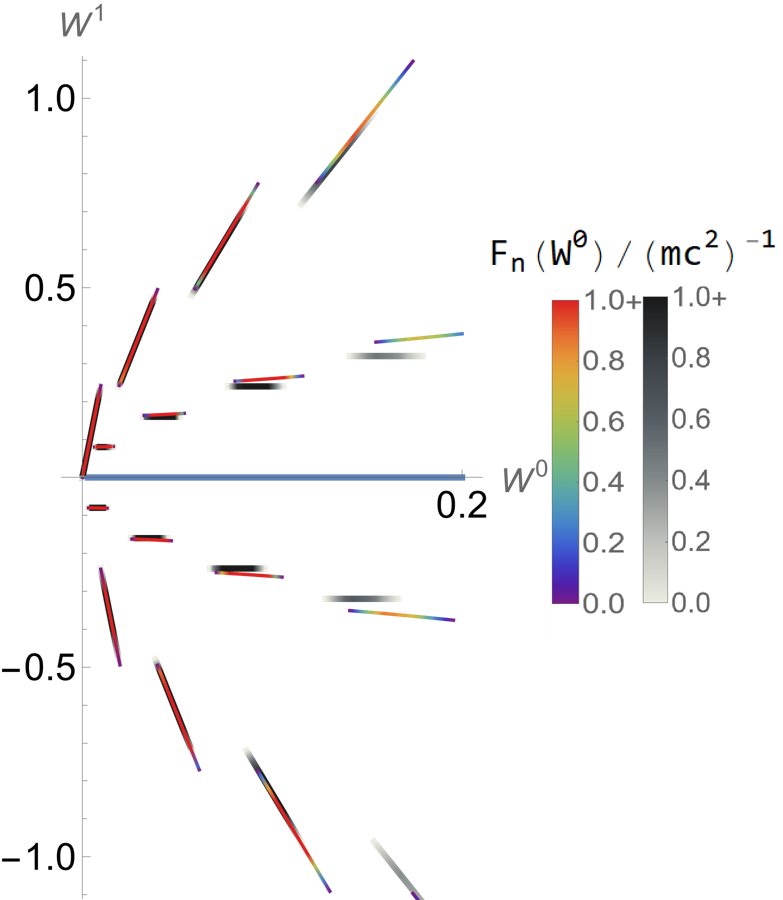}
        \caption{$Z_f/Z_i=2,v_p=0.04c,\beta=10,u=0$.}
        \label{fig:compare low T}
    \end{subfigure}
    \captionsetup{justification=raggedright,singlelinecheck=false}
    \caption{Comparison of the joint distributions of four-vector work for relativistic and non-relativistic dynamics.
    The parameters are shown below each figure.
    The thin rainbow-colored curves represent the four-vector work distribution of the relativistic piston model, while the thick curves shaded according to the GrayTones (reversed) scheme corresponds to the non-relativistic piston model. In the low-temperature and slow-driving regime, the relativistic distribution converges toward its non-relativistic counterpart.}
    \label{fig:Compare of G and L}
\end{figure}

Despite the consistency, the results from the Lorentz and Galileo covariant work distributions also display discrepancies. To identify the conditions under which the deviation is prominent, the two distributions are plotted under different driving velocity and temperature circumstances. From the results illustrated in FIG.\ref{fig:Compare of G and L}, it can be concluded that the Lorentz covariant results deviate more significantly from the Galileo covariant results at high temperature and fast speed.

\section{The covariant fluctuation theorems of work in 1D piston model}\label{section:hs}
Fluctuation theorems bridge the gap between microreversibility and macroscopic irreversibility. By adopting a step-by-step coarse-graining procedure, one is expected to derive a hierarchical structure of fluctuation theorems of work, starting from Liouville's Theorem to detailed fluctuation theorem, the Crooks fluctuation theorem and the Jarzynski equility \cite{quan23}. These FTs also hold in the relativistic piston model. In this section, we will obtain this family of covariant FTs of work, and demonstrate their validity with the results of the four-vector work distributions obtained in the previous section.

\subsection{Liouville's Theorem in 1D Piston Model}
\begin{figure}[htbp]
    \centering
    \includegraphics[width=0.6\linewidth]{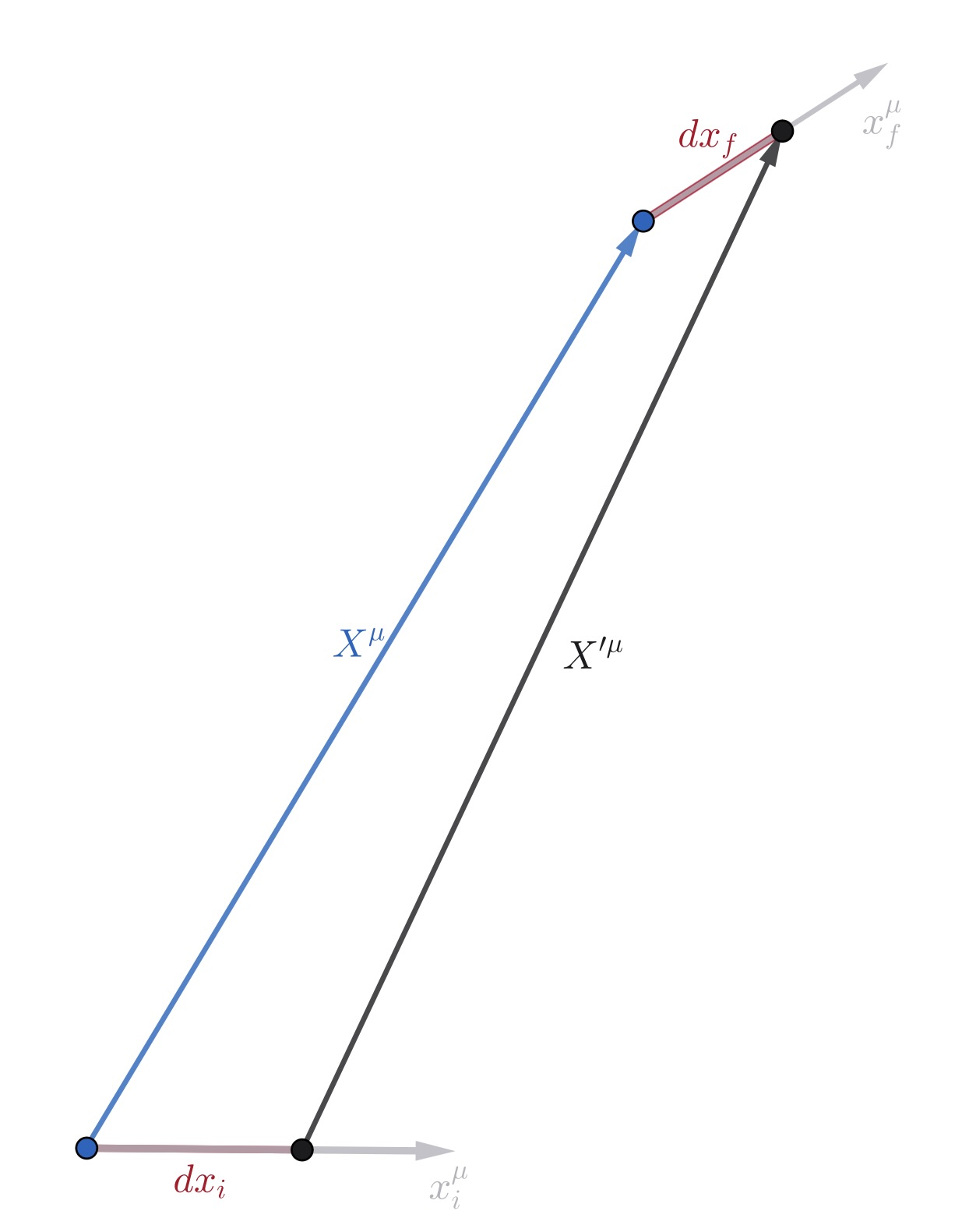}
    \captionsetup{justification=raggedright,singlelinecheck=false}
    \caption{Illustration of verifying Liouville's Theorem in the 1D piston model.}
    \label{fig:1dlt}
\end{figure}

In Sec.~\ref{sec:model}, we introduce an orthogonal reflection method which transforms the worldline of the particle into a straight auxiliary worldline.
Thus, we may discuss the Liouville's theorem for a free particle, and the results should also be valid for the piston model.
As illustrated in FIG.~\ref{fig:1dlt}, in order to verify the covariant Liouville's theorem for a free particle, we will start by perturbing the initial and final state in the Minkowski space and determine how a vector $X^{\mu}$ connecting the initial and the final spacetime coordinates will alter accordingly. 
In relativity, the measurement of a component of a Lorentz four-vector gives the inner product of the four-vector and a conjugate unit vector. In our case, in order to measure the position (space coordinate), the unit vector is chosen to be space-like. Thus, the perturbations can be denoted as vectors along $x_i^{\mu}$ and $x_f^{\mu}$ with lengths $dx_i$ and $dx_f$. The vector $X^{\mu}$ will therefore become $X'^{\mu}$.

Now, consider what happens to the four-velocity $u^{\mu}$ of the particle. $u^{\mu}$ is defined as $u^{\mu}=\frac{X^{\mu}}{\lvert X^{\mu}\rvert}\equiv\frac{X^{\mu}}{\Delta\tau}$. Denote $dX^{\mu}$ as $X'^{\mu}-X^{\mu}$. The four-velocity after the perturbation is $u^{\mu}+du^{\mu}=\frac{X^{\mu}+dX^{\mu}}{\Delta \tau+d\tau}$. The detailed of the four-velocity is:
\begin{equation}
    \begin{split}
    du^{\mu}=\frac{1}{\Delta \tau}(dX^{\mu}-u^{\mu}d\tau),
    \end{split}
\end{equation}
here $dX^{\mu}=x_f^{\mu}dx_f-x_i^{\mu}dx_i$. 
The term $d\tau$ denotes the change in the length of the timelike four-vector $X^{\mu}$. By expanding $d\tau$ to the first-order, we obtain
\begin{equation}
    \begin{split}
    d\tau=u_{\nu}x_f^{\nu}dx_f-u_{\nu}x_i^{\nu}dx_i.
    \end{split}
\end{equation}
which follows Einstein's summation convention. Thus, we have
\begin{equation}
    \begin{split}
    du^{\mu}=\frac{1}{\Delta\tau}(x_f^{\mu}dx_f-x_i^{\mu}dx_i-u^{\mu}u_{\nu}x_f^{\nu}dx_f+u^{\mu}u_{\nu}x_i^{\nu}dx_i).
    \end{split}
\end{equation}

To measure the momentum, the measurement unit vector is the same as the vector used to measure the coordinate. This measurement can be analytically expressed as $dp_i=du^{\mu}x_{i\mu}$ and $dp_f=du^{\mu}x_{f\mu}$. Thus,
\begin{equation}
    \begin{split}
        \frac{\partial p_i}{\partial x_f}=\frac{x_{i\mu}}{\Delta \tau}(x_f^{\mu}-u^{\mu}u_{\nu}x_f^{\nu});\\
        \frac{\partial p_f}{\partial x_i}=\frac{x_{f\mu}}{\Delta \tau}(-x_i^{\mu}+u^{\mu}u_{\nu}x_i^{\nu}).
    \end{split}
\end{equation}

Notice that for any contravariant vector $a_{\mu}$ and covariant vector $b^{\mu}$, $a_{\mu}b^{\mu}=b_{\mu}a^{\mu}$. Therefore, it can be concluded that $\lvert\frac{\partial p_i}{\partial x_f}\rvert=\lvert\frac{\partial p_f}{\partial x_i}\rvert$.

Consider two Jacobian determinants $\left\lvert\frac{\partial (x_i,p_i)}{\partial (x_i,x_f)}\right\rvert$ and $\left\lvert\frac{\partial (x_f,p_f)}{\partial (x_i,x_f)}\right\rvert$ :
\begin{equation}
    \begin{split}
        \left\lvert\frac{\partial (x_i,p_i)}{\partial (x_i,x_f)}\right\rvert=\begin{vmatrix}
\frac{\partial x_i}{\partial x_i} & \frac{\partial p_i}{\partial x_i} \\
\frac{\partial x_i}{\partial x_f} & \frac{\partial p_i}{\partial x_f}
\end{vmatrix}=\begin{vmatrix}
1 & \frac{\partial p_i}{\partial x_i} \\
0 & \frac{\partial p_i}{\partial x_f}
\end{vmatrix}=\frac{\partial p_i}{\partial x_f};\\
\left\lvert\frac{\partial (x_f,p_f)}{\partial (x_i,x_f)}\right\rvert=\begin{vmatrix}
\frac{\partial x_f}{\partial x_i} & \frac{\partial p_f}{\partial x_i} \\
\frac{\partial x_f}{\partial x_f} & \frac{\partial p_f}{\partial x_f}
\end{vmatrix}=\begin{vmatrix}
0 & \frac{\partial p_f}{\partial x_i} \\
1 & \frac{\partial p_f}{\partial x_f}
\end{vmatrix}=-\frac{\partial p_f}{\partial x_i}.
    \end{split}
\end{equation}

Utilizing the relation $\lvert\frac{\partial p_i}{\partial x_f}\rvert=\lvert\frac{\partial p_f}{\partial x_i}\rvert$ that we have obtained, it can be concluded that $\left\lvert\left\lvert\frac{\partial (x_i,p_i)}{\partial (x_i,x_f)}\right\rvert\right\rvert=\left\lvert\left\lvert\frac{\partial (x_f,p_f)}{\partial (x_i,x_f)}\right\rvert\right\rvert$. Hence,
\begin{equation}
    \begin{split}
        \left\lvert\frac{\partial (x_i,p_i)}{\partial (x_f,p_f)}\right\rvert=1.
    \end{split}
\end{equation}
Liouville's theorem is verified in this one-dimensional piston model.

Liouville's theorem encodes the microreversibility in the one-dimensional piston model.
For a forward trajectory $\gamma$, if we denote the reverse trajectory as $\tilde{\gamma}$, then the probability densities of the forward and the reverse trajectories satisfy
\begin{equation}
    \mathcal{P}[\gamma|\gamma(0)]=\tilde{\mathcal{P}}[\tilde\gamma|\tilde\gamma(0)],
\end{equation}
from which we can derive a family of FTs by adopting a step-by-step coarse-graining procedure \cite{quan23}.

\subsection{Covariant Detailed Fluctuation Theorem in 1D Piston Model}
As we have previously assumed, the velocity of the particle considered in the relativistic piston model follows the Maxwell-J\"uttner distribution. For a given forward trajectory $\gamma$ with a known trajectory work $W^{\mu}$, the probability density on the phase space $dxdp$ reads $\frac{1}{2K_1(\beta)L_i}e^{-\beta \cosh(y_i')}$, where $y_i'$ denotes the initial rapidity of the particle observed in the cylinder frame. The ratio of the probability density between the forward and the reverse process (denoted as $\mathcal{P}_{forward}$ and $\mathcal{P}_{reverse}$ in the following discussion) gives:
\begin{equation}
    \begin{split}
        \frac{\mathcal{P}_{forward}(\gamma(0))}{\mathcal{P}_{reverse}(\tilde{\gamma}(0))}=\frac{\exp(-\beta \cosh(y_i'))}{L_i}\cdot\frac{L_f}{\exp(-\beta \cosh(y_f'))}.
    \end{split}
\end{equation}
In which $\gamma(0)$ denotes the initial state in the phase space, while $\tilde{\gamma}(0)$ denotes the final state in the phase space. The subscript $f$ in $L_f$ and $y_f'$ stands for final state.

For the factor $\exp(-\beta \cosh(y'))$, consider the exponential term $-\beta \cosh(y')$. For an observer moving at velocity $\tanh(y_u)$ relative to the cylinder, the four-vector inverse temperature of the initial canonical ensemble perceived by the observer is $\beta_\mu=\beta u_\mu=\beta (\cosh{y_u},\sinh{y_u})$. 
If a particle has the rapidity $y'$ in the cylinder's frame, then the rapidity of the particle measured from the observer's frame is $y'-y_u$, and the four momentum is $P^\mu=(\cosh{(y'-y_u)},\sinh{(y'-y_u)})$.
The term $\beta_{\mu}P^{\mu}$ that the observer observed is
\begin{equation*}
    \beta (\cosh(y_u), \sinh{y_u})\cdot (\cosh(y'-y_u),\sinh(y'-y_u))=\beta \cosh(y').
\end{equation*}
Hence, $\exp(-\beta \cosh(y'))=\exp(-\beta_{\mu}P^{\mu})$. Utilizing this relation, we obtain $\frac{\exp(-\beta \cosh(y_i'))}{\exp(-\beta \cosh(y_f'))}=\exp(\beta_{\mu}(P_f^{\mu}-P_i^{\mu}))=\exp[\beta_{\mu}W^{\mu}(\gamma(0))]$.

The term $\frac{L_f}{L_i}$ naturally yields $\exp(-\beta \Delta F)$, and with this relation, we obtain
\begin{equation}
    \frac{\mathcal{P}_{forward}(\gamma(0))}{\mathcal{P}_{reverse}(\tilde{\gamma}(0))}=\exp(\beta_{\mu}W^{\mu}(\gamma(0))-\beta \Delta F).
\end{equation}

This is exactly the expression of detailed fluctuation theorem \cite{jihui25}. Thus, detailed fluctuation theorem is verified in this model.
\subsection{Covariant Crooks Fluctuation Theorem in 1D Piston Model}
By doing integration over all the different trajectories with the same trajectory work 
$W^{\mu}$, we obtain the Crooks fluctuation theorem 
\begin{equation}\label{eqn:crooks}
    \frac{\tilde{F}(-W^{\mu})}{F(W^{\mu})}=\exp(-\beta_{\mu}W^{\mu}+\beta \Delta F). 
\end{equation}

\begin{figure}
    \centering
    \captionsetup{justification=raggedright,singlelinecheck=false}
    \includegraphics[width=1\linewidth]{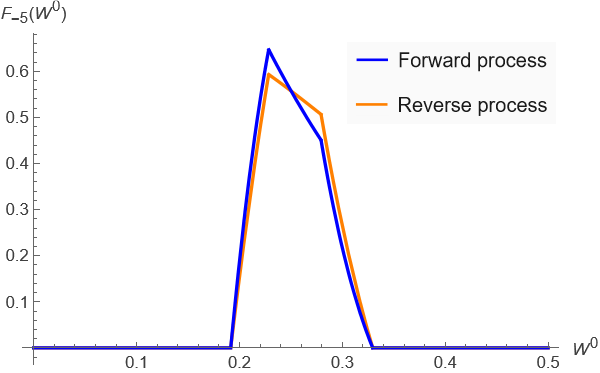}
    \caption{Intersection between the probability distribution of the forward and the backward processes. In this scenario, the collision count is -5, the ratio between the initial and final cylinder length is 2.72, $\beta=4.00$, $\Delta F=\ln(L_f/L_i)/\beta=0.25$.}
    \label{fig:non covariant crooks}
\end{figure}

Next, we consider the physical picture of covariant Crooks fluctuation theorem. For the non-covariant case, Crooks fluctuation theorem indicates that by drawing the forward and the backward processes' probability distribution as functions of the trajectory work and determine its intersection, it can be observed that the two probability distributions crosses each other at $W=\Delta F$. This relation is illustrated in the FIG.~\ref{fig:non covariant crooks}.

For the covariant case, the intersections should satisfy $u_{\mu}W^{\mu}=\Delta F$, indicating the intersections fall on a straight line in the $(W^{0}-W^{1})$ space. 
This conclusion can be observed in the following figures (see FIGs.~\ref{fig:Crooks Static}, \ref{fig:Crooks Move}). 
In FIG.~\ref{fig:Crooks Static}, for the observer at rest with the cylinder, it can be found that all the crossing points of $(W^0, W^1)$ for various collision counts locate on a straight line $W^{0}=\Delta F$ (because $u_\mu=(1,0)$ in this case).
In FIG.~\ref{fig:Crooks Move}, under the same condition except that the observer is moving at the velocity $-u=-0.15c$, the crossing points vary from the static case. Nevertheless, they all fall on the straight line $u_{\mu}W^{\mu}=\Delta F$.

\begin{figure}
    \centering
    \captionsetup{justification=raggedright,singlelinecheck=false}
    \includegraphics[width=0.5\linewidth]{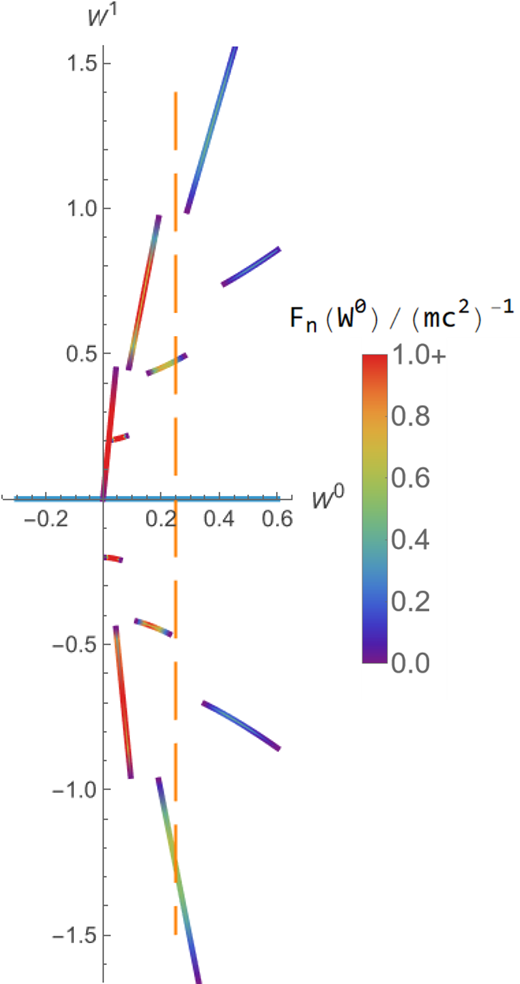}
    \caption{the Crooks fluctuation theorem, observed from the cylinder frame. By connecting the intersections between the probability distribution curves of the forward and the backward processes, we obtain the yellow dashed line shown in the graph. Here, the parameters for the forward process are chosen to be $L_i=1, L_f=2.72, v_p=0.1, \tau=10, \beta=4.00, u=0$.}
    \label{fig:Crooks Static}
\end{figure}

\begin{figure}
    \centering
    \includegraphics[width=0.5\linewidth]{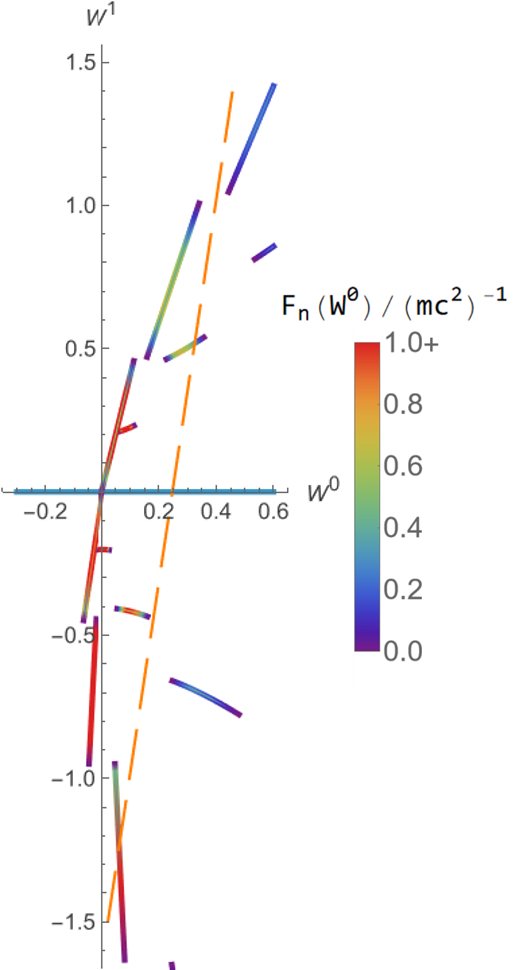}
    \captionsetup{justification=raggedright,singlelinecheck=false}
    \caption{the Crooks Fluctuation Theorem, observed from the frame with velocity $u=0.15c$. By connecting the intersections between the probability distribution curve of the forward and the backward process, we obtain the yellow dashed line shown in the graph. In this case, the yellow line follows the relation $u_{\mu}W^{\mu}=\Delta F$, instead of $W^0=\Delta F$.}
    \label{fig:Crooks Move}
\end{figure}

The previous discussions about fluctuation theorems provide deeper insights into covariant thermodynamics. While for inertial observers with different velocities, the four-vector work distributions they observed are different, the covariant fluctuation theorems remain the same in arbitraray inertial frame of reference. That is, the covariant forms of thermodynamic laws are universally valid in any inertial frame of reference.

\subsection{\label{subsec:je}Covariant Jarzynski Equality in 1D Piston Model}
The covariant JE reads \cite{jihui25}
\begin{equation}
    \langle e^{\beta_\mu W^\mu}\rangle=e^{-\beta(F_f-F_i)}=\frac{Z_f}{Z_i},
\end{equation}
here $\beta_\mu=\beta u_{\mu}$ is the four-vector inverse temperature of the initial canonical ensemble \cite{kampen68}, $u_\mu$ is the four-velocity of the cylinder, $W^\mu$ is the four-vector work which is the inverse of the change of four-momentum during the driving process (the convention in Refs.~\cite{lua05,xianghang24} and the current article define the work as the work output, which is slightly different from the typical convention \cite{jar97,jihui25}), $F_{i}$ and $F_{f}$ are the initial and final free energy of the thermal state defined on the hypersurface $\Sigma_{\mathrm{ini}}$ and $\Sigma_{\mathrm{fin}}$,
$Z_i$ and $Z_f$ are the initial and final partition function corresponding to the thermal state on $\Sigma_{\mathrm{ini}}$ and $\Sigma_{\mathrm{fin}}$.

The partition function is an integration on a space-like hypersurface.
In our setup, $Z_i$ and $Z_f$ are \cite{hakim11}
\begin{equation}
    Z_{i/f}=\int_{\Sigma_{i/f}}d\Sigma_\mu \frac{P^\mu}{m} \frac{dP^1}{P^0} \exp(-\beta_\mu P^\mu), 
\end{equation}
here $d\Sigma_\mu$ is a space-like surface element on $\Sigma_{i/f}$, $P^1$ is the momentum of the particle, $P^\mu=(\sqrt{1+{P^1}^2},P^1)$ is the four-momentum of the particle's initial state, $dP^1/P^0$ is the Lorentz invariant measure in the phase space, $\exp(-\beta_\mu P^\mu)$ is the J\"uttner-Synge factor \cite{jut11,synge57,hakim11}.
For the thermal state with the same $\beta_\mu$ on $\Sigma_i$ and $\Sigma_f$, we no longer need to compute the details of integration in momentum space.
The ratio $Z_f/Z_i$ is simply the ratio between the area $\Sigma_i$ and $\Sigma_f$, and the ratio is $(L+v_p \tau)/L$.

\begin{figure}[htbp]
    \centering
    \includegraphics[width=0.8\linewidth]{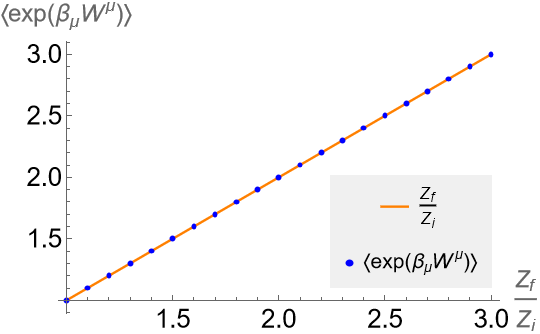}
    \captionsetup{justification=raggedright,singlelinecheck=false}
    \caption{Expectation value of the exponential four-vector work under different protocols. 
    In this figure, the driving time is changed, thus the ratio between the initial and final length of the vessel (which equals $Z_f/Z_i$) is changed. 
    The observer is moving with velocity $u=0.2c$ backwards in the $x$ direction, the inverse temperature in the rest frame of the reservoir is $\beta =5$, the velocity of the piston in the cylinder's frame is $v_p=0.1c$.
    In this numerical calculation, the error of the JE is on the same order of magnitude as the error in the normalization of probability, indicating that the covariant JE is exact.}
    \label{fig:different protocol}
\end{figure}

\begin{figure}[htbp]
    \centering
    \includegraphics[width=0.8\linewidth]{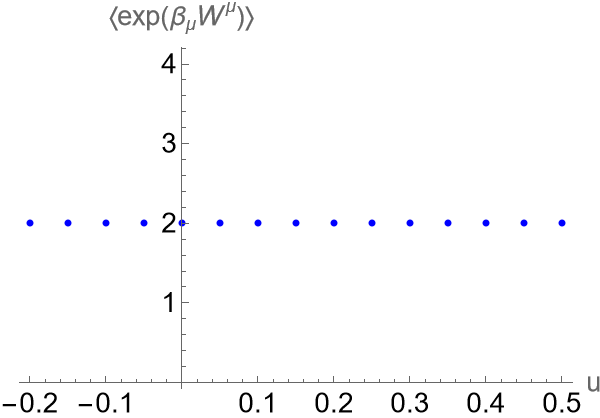}
    \captionsetup{justification=raggedright,singlelinecheck=false}
    \caption{Expectation value of the exponential four-vector work perceived from different inertial observers. 
    In this figure, the protocol is fixed, thus the ratio between the initial and final length of the vessel (which equals $Z_f/Z_i$) is fixed to be 2 (by setting $L_i=1,v_p=0.1,\tau=10$), the inverse temperature in the rest frame of the reservoir is $\beta =5$.
    The observers are moving with different $u$ backwards in the $x$ direction.
    For different observers, although the joint distributions of the four-vector work are different, the expectation value of the exponential four-vector work are exactly the same, demonstrating the validity of the covariant JE.}
    \label{fig:different observer}
\end{figure}

The covariant JE is a direct result of Eq.~(\ref{eqn:crooks}).
We can also numerically verify JE for an arbitrary moving observer with the analytical expression of the distribution of four-vector work (see FIG.\ref{fig:different protocol}, FIG.\ref{fig:different observer}).
In FIG.\ref{fig:LT of work distribution}, we have shown that the joint distributions of the four-vector work for different inertial observers are related via a Lorentz transformation.
The rapidity of each straight line segment is Lorentz boosted by the same value $y_u$, and the positions of both ends of the nonzero part on each hyperbola are also Lorentz boosted under the same principle (the hyperbolas themselves are fixed because each hyperbola indicates a constant spacetime interval, which is Lorentz invariant).
However, the exponent $\beta_\mu W^\mu$ in the covariant JE is a Lorentz scalar. 
Meanwhile, the probability measure is also a Lorentz scalar.
This explains why the expectation value of the exponential work is Lorentz invariant.

\section{Discussion and Summary\label{sec:dis}}
We derive the joint distribution of four-vector work perceived by an arbitrary inertial observer.
The first feature of the joint distribution function of four-vector work is that the probability concentrate on the origin and some curves in the $(W^0,W^1)$ space, instead of any smooth distribution.
The second feature of the distribution is that the concentration line is either linear or hyperbolic, which reflects the main features of Lorentz transformation: 
The Lorentz transformation is a linear transformation maintaining the space-time interval.
We investigate the difference between relativistic and Newtonian mechanics, and find the difference in both the magnitude and the position of the $W^0-W^1$ line.

Our model also serves as a pedagogical model in relativistic stochastic thermodynamics.
We demonstrate the validity of FTs of work, from the fine-grained covariant Liouville's theorem and detailed FT on the trajectory level, to the coarse-grained FTs, e.g., the Crooks fluctuation theorem and the JE.
We also show that the momentum component of the four-vector work contributes significantly to the average exponential work, even in the non-relativistic limit.
In addition, we illustrate that the work distributions of the forward and the backward processes cross at $u_\mu W^\mu=\Delta F$.

Technically, we developed a new method to solve the dynamics of the collision process.
The geometrical solution avoids the tedious calculation in the recursion of the time of collisions \cite{xianghang24}, providing a clear physical intuition of the relativistic collision motion.
This method could be generalized to (2+1) dimensional or (3+1) dimensional relativistic piston model directly.
For example, in the (2+1) dimensional piston model, the boundary of the vessel is a line, and in Minkowski space, the motion of the line can be described by a worldsheet-like hypersurface.
We may still introduce the orthogonal reflection for the hypersurface, which reverses the orthogonal component while maintaining the parallel component with respect to the hypersurface of the boundary.
An orthogonal reflection still characterizes an elastic collision, and we may still introduce an extra orthogonal reflection to make sure the auxiliary worldline to be a straight line.
Our future work will focus on these generalization to the more realistic models.
In 2009, R. Nolte and A. Engel studied the (3+1) dimensional model \cite{nolte09}, but the number of collisions is limited to no more than one.
Now, with our new geometrical solution, it is possible to investigate a more general protocol, in which multiple collisions are allowed.
Our method may also be helpful in solving the motion of classical or quantum Klein-Gordon field with Dirichlet boundary condition \cite{moore70, koehn12}.

In summary, we study a simple model of a piston and ideal gas in the framework of the special theory of relativity.
We calculate the four-vector work of a piston system under linear quench perceived by an arbitrary inertial observer.
Using a novel coordinate transformation technique, we obtain an analytical result of the joint distribution of four-vector work (\ref{eqn:pw1}) - (\ref{eqn:1evendelta}) and verify the covariant FTs.
With our result, it is possible to observe the deviation of relativistic work distribution (\ref{eqn:pw1}) - (\ref{eqn:1evendelta}) from the non-relativistic one (\ref{eqn:luapw1}) - (\ref{eqn:2evendelta}).
In principle, these relativistic corrections become non-negligible in the high-temperature and fast-speed limit.
However, the range of parameters where relativistic effects are observable would already be far beyond the current experimental techniques \cite{xianghang24}.
We derive a family of covariant fluctuation theorems of work, which gave us deeper insights into the essence of covariance. That is, covariant FTs hold in any inertial frame of reference.
Our results show that the momentum component in the four-vector work plays a significant role in the covariant FTs, even in the non-relativistic limit.
This model gives a clear example of the covariant FTs, thus serving the pedagogical purpose.

\section*{Acknowledgements}
Chentong Qi thanks Yichen Xiao for helpful discussions.
Tingzhang Shi thanks Yukawa Institute for Theoretical Physics at Kyoto University for the helpful discussion during the "Kyoto Workshop on Quantum Thermodynamics and Stochastic Thermodynamics 2025" (YITP-I-25-03).

This work is supported by the National Natural Science Foundations of China (NSFC) under Grants No. 12375028 and No. 12521004.

\appendix
\section{Solution of the Overlap Function}\label{appendix:overlap}
For a given number of collisions, the four-vector trajectory work does not depend on the initial position, thus when we calculate the expectation value in the state space and do the integration over the spatial component, we usually encounter the overlap function.
The overlap function is a Lorentz scalar, therefore we may apply it safely when we calculate the measure of probability.
For a moving observer, we may perform a backward Lorentz transformation to the cylinder's frame and calculate this Lorentz scalar.

In the cylinder's frame, for an even number of collisions $n=2k$, the $2k$-th image of $\Sigma_{\mathrm{fin}}$ is the line segment $C_{\mathrm{f,k}}-P_{\mathrm{f,k}}$ (see FIG.~\ref{fig:domain and overlap}).
With the coordinates of the boundary points of $\Sigma_{\mathrm{fin,n}}$ in Eq.~(\ref{eqn:coordinate rest frame}), the boundaries of the ``shadow" is denoted as the line segment $\xi_{\mathrm{C,k}} - \xi_{\mathrm{P,k}}$, with coordinates
\begin{equation}\label{eqn:shadow boundaries}
    \begin{split}
        [t(\xi_{\mathrm{C,k}}),x(\xi_{\mathrm{C,k}})]&=\left[0,x(C_{\mathrm{f,k}})-v_i (t(C_{\mathrm{f,k}})-\frac{L_i}{v_p})\right],
        \\
        [t(\xi_{\mathrm{P,k}}),x(\xi_{\mathrm{P,k}})]&=
       \left[0,x(P_{\mathrm{f,k}})-v_i (t(P_{\mathrm{f,k}})-\frac{L_i}{v_p})\right],
    \end{split}
\end{equation}
and the overlap function \cite{xianghang24} is the length of the intersection between the ``shadow" and $\Sigma_{\mathrm{ini}}$, which equals
the length $l([0, L_i]\cap [x(\xi_{\mathrm{C,k}}),x(\xi_{\mathrm{P,k}})])$, here $l(A)$ represents the length, or the measure, of set $A$.
For an odd number of collisions $n=2k-1$, the $(2k-1)$-th image of  $\Sigma_{\mathrm{fin}}$ becomes the line segment $P_{\mathrm{f,k-1}}-C_{\mathrm{f,k}}$ (see FIG.~\ref{fig:domain and overlap}).
The corresponding ``shadow" of $\Sigma_{\mathrm{fin}}$ becomes the line segment $\xi_{\mathrm{P,k-1}} - \xi_{\mathrm{C,k}}$, whose coordinates are given in Eq.~(\ref{eqn:shadow boundaries}).
The overlap function equals the length $l([0, L_i]\cap [x(\xi_{\mathrm{P,k-1}}),x(\xi_{\mathrm{C,k}})])$.

To conclude, the overlap function is
\begin{equation}
    \varphi_n(v_i)=\begin{cases}
       l([0, L_i]\cap [x(\xi_{\mathrm{C,k}}),x(\xi_{\mathrm{P,k}})]) &,n=2k\\
        l([0, L_i]\cap [x(\xi_{\mathrm{P,k-1}}),x(\xi_{\mathrm{C,k}})])&,n=2k-1.
    \end{cases}
\end{equation}
With Eqs.~(\ref{eqn:distance}), (\ref{eqn:coordinate rest frame}), and (\ref{eqn:shadow boundaries}), we derive the explicit expressions as:
\begin{widetext}
    \begin{equation}\label{eqn:overlap}
        \begin{split}
            \varphi_{2k-1}(v_i)
                =&\begin{cases}
                &L_i-S_P \sinh((2k-1)y_p) +v_i (S_P \cosh((2k-1)y_p)-\frac{L_i}{v_p}), \\
                &\qquad \qquad v_i\in \left[\frac{S_P \sinh((2k-1)y_p)-L}{S_P \cosh((2k-1)y_p)-L_i\cdot v_p{-1}}, \frac{S_P \sinh((2k-1)y_p)}{S_P \cosh((2k-1)y_p)-L_i\cdot v_p^{-1}}\right];\\
                &L_i,\\ 
                &\qquad \qquad v_i\in \left[\frac{S_P \sinh((2k-1)y_p)}{S_P \cosh((2k-1)y_p)-L_i\cdot v_p^{-1}},\frac{S_C \sinh(2ky_p)-L}{S_C \cosh(2ky_p)-L_i\cdot v_p^{-1}} \right];\\
                &S_C \sinh(2ky_p) -v_i (S_C \cosh(2ky_p)-\frac{L_i}{v_p}),\\
                &\qquad \qquad v_i\in \left[\frac{S_C \sinh(2ky_p)-L}{S_C \cosh(2ky_p)-L_i\cdot v_p^{-1}},\frac{S_C \sinh(2ky_p)}{S_C \cosh(2ky_p)-L_i\cdot v_p^{-1}}\right];
            \end{cases}\\
            \varphi_{2k}(v_i)
                =&\begin{cases}
                 &L_i-S_C \sinh(2ky_p) +v_i (S_C \cosh(2ky_p)-\frac{L_i}{v_p}),\\
                 &\qquad \qquad v_i\in \left[\frac{S_C \sinh(2ky_p)-L}{S_C \cosh(2ky_p)-L_i\cdot v_p^{-1}}, \frac{S_C \sinh(2ky_p)}{S_C \cosh(2ky_p)-L_i\cdot v_p^{-1}}\right];\\
                &L_i, \\
                &\qquad \qquad v_i\in \left[\frac{S_C \sinh(2ky_p)}{S_C \cosh(2ky_p)-L_i\cdot v_p^{-1}}, \frac{S_P \sinh((2k+1)y_p)-L}{S_P \cosh((2k+1)y_p)-L_i\cdot v_p^{-1}}\right];\\
                &S_P \sinh((2k+1)y_p) -v_i (S_P \cosh((2k+1)y_p)-\frac{L_i}{v_p}),\\
                &\qquad \qquad v_i\in \left[\frac{S_P \sinh((2k+1)y_p)-L}{S_P \cosh((2k+1)y_p)-L_i\cdot v_p^{-1}}, \frac{S_P \sinh((2k+1)y_p)}{S_P \cosh((2k+1)y_p)-L_i\cdot v_p^{-1}}\right].
            \end{cases}
        \end{split}
    \end{equation}
\end{widetext}
For a moving observer, we only need to replace the initial velocity $v_i$ with $\tanh(\mathrm{arctanh}(v_i)-\mathrm{arctanh}(u))$, which is the initial velocity of the same particle, but measured in the cylinder's frame.

\section{Exact Result of Lorentz Covariant work distributions}\label{appendix:er}
As shown in Eq.(\ref{eqn:L}), deriving the exact four-vector work distribution requires integrating over a $\delta$ function, and in this case, we integrate out the $W^0_{\tau}$ term first. Notice that for $n$ of different parities, the four-vector work differs from one to another, hence it is necessary to treat odd $n$ and even $n$ separately.

In the following discussion, for simplicity, remark

\begin{equation}\label{eqn:P Lorentz}
    \begin{split}
        F_L(W^0,W^1)=\sum_nF_n(W^0)\delta\left(W^1-W^1_{\tau}(W^0)\right),
    \end{split}
\end{equation}
where
\begin{widetext}
\begin{equation}
    \begin{split}
    F_n(W^0)=\int_{D_n}\mathrm{d}y_i\frac{\exp\left(-\beta \cosh(y_i-y_u)\right)\cosh(y_i-y_u)}{2K_1(\beta)}\varphi_n\left(\tanh(y_i-y_u)\right)\delta\left(W^0 - W^0_{\tau}(x, y_i)\right).
    \end{split}
\end{equation}

For $n$ as an odd number (let $n=2k-1$ where $k$ is an integer):
\begin{equation}\label{eqn:lpodd}
    \begin{split}
F_{2k-1}(W^0)=\frac{\varphi_{2k-1}\left(\tanh(y_i-y_u)\right)\cosh(y_i-y_u)\exp\left(-\beta \cosh(y_i-y_u)\right)}{4K_1(\beta)\sinh(ky_p+y_u)\cosh(y_i-y_u+ky_p)};
    \end{split}
\end{equation}
\begin{equation}\label{eqn:ldodd}
    \begin{split}
        W^1_{\tau}=2\cosh(ky_p+y_u)\sinh(y_i-y_u-ky_p).
    \end{split}
\end{equation}
\end{widetext}
Seen in Eq. (\ref{eqn:wzeroy}), $y_i$ can be explicitly expressed by $W^0$ :
\begin{equation}\label{eqn:lyodd}
    \begin{split}
    y_i=y_u+ky_p+\mathrm{arcsinh}\left(\frac{W^0}{2\sinh(ky_p+y_u)}\right).
    \end{split}
\end{equation}
By substituting $y_i$ in the covariant work distribution, the explicit expression of this distribution can be obtained. The exact results are shown in Eq. (\ref{eqn:pw1}) and Eq. (\ref{eqn:1odddelta}). It can be concluded from Eq. (\ref{eqn:1odddelta}) that the work distribution lies solely on various straight lines on the $(W^0,W^1)$ space for odd $n$'s.

\begin{widetext}
    \begin{equation}\label{eqn:pw1}
        \begin{split}
        &F_{2k-1}(W^0)=\varphi_{2k-1}\left[\left(\sinh(ky_p)\sqrt{1+\left(\frac{W^0}{2\sinh(ky_p)}\right)^2}+\frac{\cosh(ky_p)W^0}{2\sinh(ky_p+y_u)}\right)\right.\times\\
        &\left.\left(\cosh(ky_p)\sqrt{1+\left(\frac{W^0}{2\sinh(ky_p+y_u)}\right)^2}+\frac{W^0\sinh(ky_p)}{2\sinh(y_u+ky_p)}\right)^{-1}\right] \cdot\left(4K_1(\beta)\sinh(ky_p+y_u)\sqrt{1+\left(\frac{W^0}{2\sinh(ky_p+y_u)}\right)^2}\right)^{-1}
       \cdot \\
        &\exp\left[-\beta\left(\cosh(ky_p)\sqrt{1+\left(\frac{W^0}{2\sinh(ky_p+y_u)}\right)^2}+\frac{\sinh(ky_p)W^0}{2\sinh(ky_p+y_u)}\right)\right]\cdot\\
        &\left(\cosh(ky_p)\sqrt{1+\left(\frac{W^0}{2\sinh(ky_p+y_u)}\right)^2}+\frac{\sinh(ky_p)W^0}{2\sinh(ky_p+y_u)}\right);
        \end{split}
    \end{equation}
    \begin{equation}\label{eqn:1odddelta}
    \begin{split}
        \delta\left(W^1-W^1_{\tau}(W^0)\right)=\delta\left(W^1-\frac{\cosh(y_u+ky_p)W^0}{\sinh(y_u+ky_p)}\right).
        \end{split}
    \end{equation}
\end{widetext}

For $n$ as an even number (let $n=2k$ where $k$ is an integer):
\begin{widetext}
\begin{equation}\label{eqn:lpeven}
    \begin{split}
    F_{2k}(W^0)=\frac{\varphi_{2k}\left(\tanh(y_i-y_u)\right)\cosh(y_i-y_u)\exp\left(-\beta \cosh(y_i-y_u)\right)}{4K_1(\beta)\cosh(y_i-ky_p)\sinh(ky_p)}.
    \end{split}
\end{equation}
\begin{equation}\label{eqn:ldeven}
    \begin{split}
        W^1_{\tau}=2\cosh(y_i-ky_p)\sinh(ky_p).
    \end{split}
\end{equation}
\end{widetext}
Repeat the previous procedure, $y_i$ can be expressed as a function of $W^0$:
\begin{equation}\label{eqn:lyeven}
    \begin{split}
    y_i=\mathrm{arcsinh}\left(\frac{W^0}{2\sinh(ky_p)}\right)+ky_p,
    \end{split}
\end{equation}
which yields the probability distribution expression (the detailed results are listed in Eq. (\ref{eqn:pw2}) and Eq. (\ref{eqn:1evendelta})). It can be concluded from Eq. (\ref{eqn:1evendelta}) that the work distribution lies solely on various hyperbolas on the $(W^0,W^1)$ space for even $n$'s.

\begin{widetext}
    \begin{equation}\label{eqn:pw2}
        \begin{split}
        &F_{2k}(W^0)=\varphi_{2k}\left[\left(\sinh(ky_p-y_u)\sqrt{1+\left(\frac{W^0}{2\sinh(ky_p)}\right)^2}+\frac{W^0\cosh(ky_p-y_u)}{2\sinh(ky_p)}\right)\right.\times\\
        &\left.\left(\cosh(ky_p-y_u)\sqrt{1+\left(\frac{W^0}{2\sinh(ky_p)}\right)^2}+\frac{W^0\sinh(ky_p-y_u)}{2\sinh(ky_p)}\right)^{-1}\right]\cdot\\
        &\left(-\frac{W^0\sinh(y_u-ky_p)}{2\sinh(ky_p)}+\sqrt{1+\left(\frac{W^0}{2\sinh(ky_p)}\right)^2}\cosh(y_u-ky_p)\right)\cdot\left(4K_1(\beta)\sqrt{1+\left(\frac{W^0}{2\sinh(ky_p)}\right)^2}\sinh(ky_p)\right)^{-1}\cdot \\
        &\exp\left[-\beta \left(-\frac{W^0\sinh(y_u-ky_p)}{2\sinh(ky_p)}+\sqrt{1+\left(\frac{W^0}{2\sinh(ky_p)}\right)^2}\cosh(y_u-ky_p)\right)\right];
        \end{split}
    \end{equation}
\begin{equation}\label{eqn:1evendelta}
    \begin{split}
        \delta\left(W^1-W^1_{\tau}(W^0)\right)=\delta \left(W^1-2\sqrt{1+\left(\frac{W^0}{2\sinh(ky_p)}\right)^2}\sinh(ky_p)\right).
    \end{split}
\end{equation}
\end{widetext}

\section{Exact Result of Galileo covariant Work Distributions}\label{appendix:g}
In the case of Galileo covariance, for $n=2k-1$ where $k$ is an integer, by replacing $v_i$ in Eqs. (\ref{eqn:gpodd}) and (\ref{eqn:gdodd}) with Eq. (\ref{gvodd}), the distribution reads:
\begin{widetext}
\begin{equation}\label{eqn:luapw1}
    \begin{split}
        F_{2k-1}(W^0)=\frac{\sqrt{\beta}}{\sqrt{2\pi}L_i}\varphi_{2k-1}\left(\frac{W^0+2\left((k-1)u-kw\right)^2}{2(kw-(k-1)u)}\right)\cdot
        \exp\left[-\frac{\beta}{2}\left(\frac{W^0+2((k-1)u-kw)^2}{2(kw-(k-1)u)}-u\right)^2\right]\cdot\left[2(kw-(k-1)u)\right]^{-1};
    \end{split}
\end{equation}
\begin{equation}\label{eqn:2odddelta}
    \begin{split}
        \delta\left(W^1-W^1_{\tau}(W^0)\right)=\delta\left(W^1-\frac{W^0}{2\left(kw-(k-1)u\right)}\right).
    \end{split}
\end{equation}
\end{widetext}

For $n=2k$ where $k$ is an integer, by repeating the above process, the distribution reads:
\begin{widetext}
\begin{equation}\label{eqn:luapw2}
    \begin{split}
        F_{2k}(W^0)=\frac{\sqrt{\beta}}{\sqrt{2\pi}L_i}\varphi_{2k}\left(\frac{W^0+2k^2(u-w)^2}{2k(w-u)}\right)\cdot
        \exp\left[-\frac{\beta}{2}\left(\frac{W^0+2k^2(u-w)^2}{2k(w-u)}-u\right)^2\right]\cdot\left[2k(w-u)\right]^{-1};
    \end{split}
\end{equation}
\begin{equation}\label{eqn:2evendelta}
    \begin{split}
        \delta(W^1-W^1_{\tau}(W^0))=\delta(W^1-2k(w-u)).
    \end{split}
\end{equation}
\end{widetext}

\section{Details of deriving the consistency between Lorentz and Galileo covariant work distributions}\label{appendix:d}
The demonstration will be carried out in the low speed limit. We will start by considering several approximations. Because $\sinh x$$\approx x\approx$$\tanh x$ for $|x|<<1$, rapidity and velocity can be considered equivalent. As $\beta$ approaches $\infty$, the Bessel function can be approximated as ${K_1(\beta)}\approx e^{-\beta} \pi^{1/2}(2\beta)^{-1/2}$. These approximations bring an extra $\beta$ to the exponent in the Lorentz covariant four-vector work distribution.

First, consider $F_{2k-1}$ and its corresponding $\delta$ function in the Lorentz covariant case. The Lorentz covariant distribution can be divided into three segments: the exponential part, the $\delta$ function part and the rest multiplications. Taking into account the approximation of the Bessel function, the exponential part can be approximated in the following way:
\begin{widetext}
 \begin{equation}\label{eqn:expodd}
    \begin{split}
     &\exp\left[-\beta\left(\cosh(ky_p)\sqrt{1+\left(\frac{W^0}{2\sinh(ky_p+y_u)}\right)^2}+\sinh(ky_p)\frac{W^0}{2\sinh(ky_p+y_u)}\right)+\beta\right]\\
     \approx &\exp\left[\beta\left(\left(1+\frac{k^2y^2_p}{2}\right)\left(1+\frac{1}{2}\left(\frac{W^0}{2(ky_p+y_u)}\right)^2+\frac{ky_pW^0}{2(ky_p+y_u)}-1\right)\right)\right]\\
      \approx &\exp\left[-\frac{\beta}{2}\left(\frac{W^0}{2(u+kw-ku)}+k(w-u)\right)^2\right]
      =\exp\left[-\frac{\beta}{2}\left(\frac{W^0+2((k-1)u-kw)^2}{2(kw-(k-1)u)}-u\right)^2\right].
      \end{split}
\end{equation}
\end{widetext}

For the $\delta$ function segment, 
\begin{equation}\label{eqn:odddelta}
    \frac{W^0}{\tanh(y_u+ky_p)}\approx\frac{W^0}{kw-(k-1)u}.
\end{equation}

For the rest of the multipliers, expand the remaining multipliers in the Lorentz covariant distribution to the first order. Notice that $v$ and $y$ are all first order small quantities, while $W^0$ is approximately proportional to $v^2$, making it a second order small quantity. The expansion leads to:
\begin{widetext}
\begin{equation}\label{eqn:oddab}
    \begin{split}
        &\left(\cosh(ky_p)\sqrt{1+\left(\frac{W^0}{2\sinh(y_u+ky_p)}\right)^2}+\frac{\sinh(ky_p)W^0}{2\sinh(ky_p+y_u)}\right)\cdot
        \left(2\sinh(ky_p+y_u)\sqrt{1+\left(\frac{W^0}{2\sinh(y_u+ky_p)}\right)^2}\right)^{-1}\\
        \approx&\left[\left(1+\frac{k^2y_p^2}{2}\right)\left(1+\frac{1}{2}\left(\frac{W^0}{2(y_u+ky_p)}\right)^2\right)+\frac{ky_pW^0}{2(ky_p+y_u)}\right]\left(1-\frac{1}{2}\left(\frac{W^0}{2(kw-(k-1)u)}\right)^2\right)\cdot\frac{1}{2(kw-(k-1)u)}\\
        \approx&\left(1+\frac{1}{2}k^2y^2_p+\frac{ky_pW^0}{2(ky_p+y_n)}\right)\cdot
       \frac{1}{2(kw-(k-1)u)}\\
       \approx&\left(1+O(v_i)\right)\cdot\frac{1}{2(kw-(k-1)u)}
       \approx\frac{1}{2(kw-(k-1)u)}.
    \end{split}
\end{equation}
\end{widetext}

Multiplying the results yielded in the previous calculations (\ref{eqn:expodd}--\ref{eqn:oddab}), it can be concluded that by taking the low speed limit, the Lorentz covariant four-vector work distribution is equivalent to the Galileo covariant four-vector work distribution when the collision count is an odd integer, see Eqs.~(\ref{eqn:pw1},\ref{eqn:1odddelta},\ref{eqn:luapw1},\ref{eqn:2odddelta}).

Then, let us consider $F_{2k}$ and its corresponding $\delta$ function in the Lorentz covariant case. The three-segments division is still valid in the following discussion. Taking into account the approximation of Bessel function, the exponential part can be approximated in the following way:
\begin{widetext}
\begin{equation}\label{eqn:evenexp}
    \begin{split}
    &\exp\left[-\beta \left(-\frac{W^0\sinh(y_u-ky_p)}{2\sinh(ky_p)}+\sqrt{1+\left(\frac{W^0}{2\sinh(ky_p)}\right)^2}\cosh(y_u-ky_p)\right)+\beta\right]\\
    \approx&
    \exp\left[-\beta\left(-\frac{W^0(u-k(w-u))}{2k(w-u)}+\left(1+\frac{1}{2}\left(\frac{W^0}{2k(w-u)}\right)^2\right)\left(1+\frac{(u-k(w-u))^2}{2}\right)-1\right)\right]\\
    \approx &\exp\left[-\frac{\beta}{2}\left(\frac{W^0}{2k(w-u)}+kw-(k+1)u\right)^2\right]
    =\exp\left[-\frac{\beta}{2}\left(\frac{W^0+2k^2(w-u)^2}{2k(w-u)}-u\right)^2\right].
    \end{split}
\end{equation}

For the $\delta$ function segment,
\begin{equation}\label{eqn:evendelta}
    \begin{split}
        2\sqrt{1+\left(\frac{W^0}{2\sinh(ky_p)}\right)^2}\sinh(ky_p)
        \approx 2ky_p(1+O(v_i)) \approx 2ky_p
        \approx 2k(w-u).
    \end{split}
\end{equation}
For the rest of the multipliers, expanding them to the first order leads to:

    \begin{equation}\label{eqn:evenab}
        \begin{split}
            &\left[-\frac{W^0\sinh(y_u-ky_p)}{2\sinh(ky_p)}+\sqrt{1+\left(\frac{W^0}{2\sinh(ky_p)}\right)^2}\cosh(y_u-ky_p)
            \right]\cdot\left(2\sqrt{1+\left(\frac{W^0}{2\sinh(ky_p)}\right)^2}\sinh(ky_p)\right)^{-1}\\
            \approx &\left(1-\frac{1}{2}\left(\frac{W^0}{2ky_p}\right)^2\right)\left[-\frac{W^0(y_u-ky_p)}{2ky_p}+\left(1+\frac{1}{2}\left(\frac{W^0}{2ky_p}\right)^2\right)\left(1+\frac{(y_u-ky_p)^2}{2}\right)\right]\cdot\frac{1}{2ky_p}\\
            \approx&\left(1+\frac{(y_u-ky_p)^2}{2}-\frac{W^0(y_u-ky_p)}{2ky_p}\right)\cdot\frac{1}{2ky_p}\\
            \approx&(1+O(v_i))\cdot\frac{1}{2ky_p}\approx\frac{1}{2k(w-u)}.
        \end{split}
    \end{equation}
\end{widetext}

Multiplying the results yielded in the previous calculations(\ref{eqn:evenexp}--\ref{eqn:evenab}), it can be concluded that by taking the low speed limit, the Lorentz covariant four-vector work distribution is equivalent to the Galileo covariant four-vector work distribution when the collision count is an even integer, see Eqs.~(\ref{eqn:pw2},\ref{eqn:1evendelta},\ref{eqn:luapw2},\ref{eqn:2evendelta}).

Thus, from the discussion above, the consistency between Lorentz covariance and Galileo covariance in the low speed limit can be verified.

\vspace{7em}

\bibliography{relativistic_piston}

@article{hanggi09,
  author  = {Dunkel, J{\"o}rn and H{\"a}nggi, Peter and Hilbert, Stefan},
  title   = {Non-local observables and lightcone-averaging in relativistic thermodynamics},
  journal = {Nat. Phys.},
  year    = {2009},
  volume  = {5},
  number  = {10},
  pages   = {741--747},
  doi     = {10.1038/nphys1395},
  url     = {https://doi.org/10.1038/nphys1395},
  issn    = {1745-2481}
}

@article{jihui25,
  title = {Promoting Fluctuation Theorems into Covariant Forms},
  author = {Pei, Ji-Hui and Chen, Jin-Fu and Quan, H. T.},
  journal = {Phys. Rev. Lett.},
  volume = {134},
  issue = {23},
  pages = {237102},
  numpages = {7},
  year = {2025},
  month = {Jun},
  publisher = {American Physical Society},
  doi = {10.1103/xlmq-g6m5},
  url = {https://link.aps.org/doi/10.1103/xlmq-g6m5}
}

@article{quan23,
  title = {Hierarchical structure of fluctuation theorems for a driven system in contact with multiple heat reservoirs},
  author = {Chen, Jin-Fu and Quan, H. T.},
  journal = {Phys. Rev. E},
  volume = {107},
  issue = {2},
  pages = {024135},
  numpages = {14},
  year = {2023},
  month = {Feb},
  publisher = {American Physical Society},
  doi = {10.1103/PhysRevE.107.024135},
  url = {https://link.aps.org/doi/10.1103/PhysRevE.107.024135}
}

@article{jar97,
  title = {Nonequilibrium Equality for Free Energy Differences},
  author = {Jarzynski, C.},
  journal = {Phys. Rev. Lett.},
  volume = {78},
  issue = {14},
  pages = {2690--2693},
  numpages = {0},
  year = {1997},
  month = {Apr},
  publisher = {American Physical Society},
  doi = {10.1103/PhysRevLett.78.2690},
  url = {https://link.aps.org/doi/10.1103/PhysRevLett.78.2690}
}

@article{crooks99,
  title = {Entropy production fluctuation theorem and the nonequilibrium work relation for free energy differences},
  author = {Crooks, Gavin E.},
  journal = {Phys. Rev. E},
  volume = {60},
  issue = {3},
  pages = {2721--2726},
  numpages = {0},
  year = {1999},
  month = {Sep},
  publisher = {American Physical Society},
  doi = {10.1103/PhysRevE.60.2721},
  url = {https://link.aps.org/doi/10.1103/PhysRevE.60.2721}
}

@article{martin08,
author = {Maragakis, Paul and Spichty, Martin and Karplus, Martin},
title = {A Differential Fluctuation Theorem},
journal = {J. Phys. Chem. B},
volume = {112},
number = {19},
pages = {6168-6174},
year = {2008},
doi = {10.1021/jp077037r},
    note ={PMID: 18331019},
URL = {https://doi.org/10.1021/jp077037r},
eprint = {https://doi.org/10.1021/jp077037r}
}

@article{parrondo07,
  title = {Dissipation: The Phase-Space Perspective},
  author = {Kawai, R. and Parrondo, J. M. R. and den Broeck, C. Van},
  journal = {Phys. Rev. Lett.},
  volume = {98},
  issue = {8},
  pages = {080602},
  numpages = {4},
  year = {2007},
  month = {Feb},
  publisher = {American Physical Society},
  doi = {10.1103/PhysRevLett.98.080602},
  url = {https://link.aps.org/doi/10.1103/PhysRevLett.98.080602}
}

@article{zhongping14,
  title = {Interference of identical particles and the quantum work distribution},
  author = {Gong, Zongping and Deffner, Sebastian and Quan, H. T.},
  journal = {Phys. Rev. E},
  volume = {90},
  issue = {6},
  pages = {062121},
  numpages = {15},
  year = {2014},
  month = {Dec},
  publisher = {American Physical Society},
  doi = {10.1103/PhysRevE.90.062121},
  url = {https://link.aps.org/doi/10.1103/PhysRevE.90.062121}
}

@ARTICLE{jut11,
       author = {{J{\"u}ttner}, Ferencz},
        title = "{Das Maxwellsche Gesetz der Geschwindigkeitsverteilung in der Relativtheorie}",
      journal = {Ann. Phys.},
         year = 1911,
        month = jan,
       volume = {339},
       number = {5},
        pages = {856-882},
          doi = {10.1002/andp.19113390503},
       adsurl = {https://ui.adsabs.harvard.edu/abs/1911AnP...339..856J}
}

@book{synge57,
    Author = {Synge, J. L.},
    Mrclass = {83.0X},
    Mrnumber = {0088362},
    Mrreviewer = {A. H. Taub},
    Publisher = {North-Holland Publishing Company, Amsterdam; Interscience Publishers Inc., New York},
    Title = {The relativistic gas},
    Year = {1957}
}

@article{koehn12,
year = {2013},
month = {jan},
publisher = {EDP Sciences, IOP Publishing and Società Italiana di Fisica},
volume = {100},
number = {6},
pages = {60008},
author = {Koehn, M.},
title = {Solutions of the Klein-Gordon equation in an infinite square-well potential with a moving wall},
journal = {Europhys. Lett.}
}

@article{moore70,
    author = {Moore, Gerald T.},
    title = {Quantum Theory of the Electromagnetic Field in a Variable-Length One-Dimensional Cavity},
    journal = {J. Math. Phys.},
    volume = {11},
    number = {9},
    pages = {2679-2691},
    year = {1970},
    month = {09}
}

@book{hakim11,
  title = {Introduction to Relativistic Statistical Mechanics: Classical and Quantum},
  ISBN = {9789814322454},
  url = {http://dx.doi.org/10.1142/7881},
  DOI = {10.1142/7881},
  publisher = {WORLD SCIENTIFIC},
  author = {Hakim,  Rémi},
  year = {2011},
  month = mar 
}

@article{lua05,
author = {Lua, Rhonald C. and Grosberg, Alexander Y.},
title = {Practical Applicability of the Jarzynski Relation in Statistical Mechanics: A Pedagogical Example},
journal = {J. Phys. Chem. B},
volume = {109},
number = {14},
pages = {6805-6811},
year = {2005},
doi = {10.1021/jp0455428}
}

@article{quan12,
  title = {Validity of nonequilibrium work relations for the rapidly expanding quantum piston},
  author = {Quan, H. T. and Jarzynski, Christopher},
  journal = {Phys. Rev. E},
  volume = {85},
  issue = {3},
  pages = {031102},
  numpages = {8},
  year = {2012},
  month = {Mar},
  publisher = {American Physical Society},
  doi = {10.1103/PhysRevE.85.031102},
  url = {https://link.aps.org/doi/10.1103/PhysRevE.85.031102}
}

@article{gong16,
  title = {Stochastic Thermodynamics of a Particle in a Box},
  author = {Gong, Zongping and Lan, Yueheng and Quan, H. T.},
  journal = {Phys. Rev. Lett.},
  volume = {117},
  issue = {18},
  pages = {180603},
  numpages = {6},
  year = {2016},
  month = {Oct},
  publisher = {American Physical Society},
  doi = {10.1103/PhysRevLett.117.180603},
  url = {https://link.aps.org/doi/10.1103/PhysRevLett.117.180603}
}

@article{fei19,
  title = {Quantum work distributions associated with the dynamical Casimir effect},
  author = {Fei, Zhaoyu and Zhang, Jingning and Pan, Rui and Qiu, Tian and Quan, H. T.},
  journal = {Phys. Rev. A},
  volume = {99},
  issue = {5},
  pages = {052508},
  numpages = {12},
  year = {2019},
  month = {May},
  publisher = {American Physical Society},
  doi = {10.1103/PhysRevA.99.052508},
  url = {https://link.aps.org/doi/10.1103/PhysRevA.99.052508}
}

@article{qiu20,
  title = {Path-integral approach to the calculation of the characteristic function of work},
  author = {Qiu, Tian and Fei, Zhaoyu and Pan, Rui and Quan, H. T.},
  journal = {Phys. Rev. E},
  volume = {101},
  issue = {3},
  pages = {032111},
  numpages = {12},
  year = {2020},
  month = {Mar},
  publisher = {American Physical Society},
  doi = {10.1103/PhysRevE.101.032111},
  url = {https://link.aps.org/doi/10.1103/PhysRevE.101.032111}
}

@article{xianghang24,
  title = {Exact work distribution and Jarzynski's equality of a relativistic particle in an expanding piston},
  author = {Zhang, Xianghang and Shi, Tingzhang and Quan, H. T.},
  journal = {Phys. Rev. E},
  volume = {110},
  issue = {2},
  pages = {024128},
  numpages = {10},
  year = {2024},
  month = {Aug},
  publisher = {American Physical Society},
  doi = {10.1103/PhysRevE.110.024128},
  url = {https://link.aps.org/doi/10.1103/PhysRevE.110.024128}
}

@article{kampen68,
    author = {van Kampen, N. G.},
    title = "Relativistic Thermodynamics of Moving Systems",
    journal = "Phys. Rev.",
    year = "1968",
    volume = "\textbf{173}",
    pages = "295--301"
}

@article{kampen69,
    author = {van Kampen, N. G.},
    title = "Relativistic Thermodynamics",
    journal = "J. Phys. Soc. Jpn. Suppl.",
    volume = "\textbf{26}",
    year = "1969",
    pages = "316--321"
}

@article{israel79,
title = {Transient relativistic thermodynamics and kinetic theory},
journal = {Ann. Phys.},
volume = {118},
number = {2},
pages = {341--372},
year = {1979},
issn = {0003-4916},
doi = {https://doi.org/10.1016/0003-4916(79)90130-1},
url = {https://www.sciencedirect.com/science/article/pii/0003491679901301},
author = {W. Israel and J.M. Stewart}
}

@article{israel81,
title = {Thermodynamics of relativistic systems},
journal = {Physica A},
volume = {106},
number = {1},
pages = {204--214},
year = {1981},
issn = {0378-4371},
doi = {https://doi.org/10.1016/0378-4371(81)90220-X},
url = {https://www.sciencedirect.com/science/article/pii/037843718190220X},
author = {W. Israel}
}

@article{nolte09,
title = {Jarzynski equation for the expansion of a relativistic gas and black-body radiation},
journal = {Physica A},
volume = {388},
number = {18},
pages = {3752-3758},
year = {2009},
issn = {0378-4371},
doi = {https://doi.org/10.1016/j.physa.2009.05.041},
url = {https://www.sciencedirect.com/science/article/pii/S0378437109004270},
author = {Roman Nolte and Andreas Engel}
}

\end{document}